\documentclass[aps,pre,longbilbiography,twocolumn]{revtex4-1}
\usepackage{amsfonts}
\usepackage{amsmath,latexsym}
\usepackage{psfrag}
\usepackage{amssymb,amsmath,amsthm}
\usepackage[dvips]{graphicx}
\usepackage{verbatim} %
\usepackage{color}
\usepackage{xcolor}
\usepackage{braket}

\newcommand{\beq}{\begin{equation}}
\newcommand{\eeq}{\end{equation}}
\newcommand{\bea}{\begin{eqnarray}}
\newcommand{\eea}{\end{eqnarray}}

\usepackage{bm}
\usepackage{color}
\usepackage{fancyhdr}

\newcommand{\bma}{\left(\begin{matrix}}
\newcommand{\ema}{\end{matrix}\right)}

\def\build#1_#2{\mathrel{\mathop{#1}\limits_{#2}}}

\definecolor{pink}{rgb}{1,0.5,0.5}
\definecolor{violet}{rgb}{1,0,1} 
\definecolor{red}{rgb}{1,0,0}
\definecolor{yellow}{rgb}{0.7,1,0}
\definecolor{orange}{rgb}{1,0.5,0}
\definecolor{white}{rgb}{1,1,1}
\definecolor{blue}{rgb}{0,0,1}
\definecolor{cyan}{rgb}{0,1,1}

\def\S{{\mathcal S}}

\def\T{{\mathcal T}}

\def\b{{\bf b}}

\def\k{{\bf k}}

\def\x{{\bf x}}
\def\y{{\bf y}}

\def\u{{\bf u}}

\def\e{\bm{ \varepsilon}}

\begin{document}


\begin{abstract}
{A canonical quantization procedure is applied to the interaction of elastic waves---phonons---with infinitely long dislocations that can oscillate about an equilibrium, straight line, configuration.} {The interaction is implemented through the well-known Peach-Koehler force.}
{For small dislocation excursions away from the equilibrium position, the quantum theory can be solved to all orders in the coupling constant.	}
We study in detail the quantum excitations of the dislocation line, and its interactions with phonons. {The consequences for the drag on a dislocation caused by the phonon wind are pointed out}. We compute the cross-section for phonons incident on the dislocation lines for an arbitrary angle of incidence.  {The consequences for thermal transport are explored, and we compare our results, involving a dynamic dislocation, with those of Klemens and Carruthers, involving a static dislocation.  In our case, the relaxation time is inversely proportional to frequency, rather  {than} directly proportional to frequency. As a consequence, the thermal transport anisotropy generated on a material by the presence of a highly-oriented array of dislocations is considerably more sensitive {to the frequency of each propagating mode, and therefore, to the temperature of the material. }}
\end{abstract}

\title{The scattering of phonons by infinitely long quantum dislocations segments and the {generation of} thermal transport anisotropy {in} a solid threaded by many parallel dislocations}

\author{Fernando Lund$^{1,2}$ and Bruno Scheihing-Hitschfeld$^{1,3}$}

\affiliation{\mbox{$^1$Departamento de F\'\i sica and $^2$CIMAT, Facultad de Ciencias
F\'\i sicas y Matem\'aticas, Universidad de Chile, Santiago, Chile} 
\mbox{$^3$Center for Theoretical Physics, Massachusetts Institute of Technology, Cambridge, MA 02139 USA} }

\date{ \today}

\maketitle
\fancyfoot{}
\thispagestyle{fancy}
\rhead{MIT-CTP/5225}
\renewcommand{\headrulewidth}{0pt}

\section{Introduction} 
{The search for efficient ways to transform waste heat into usable power has spurred  research, both basic and applied, into thermoelectric materials wherein, for example, a temperature gradient generates an electric current. These phenomena involve the transport of energy and of electric charge, and are dominated by electrons and lattice vibrations: electrons carry both electric charge and energy, while phonons carry energy. The desire is then to optimize electron mobility (to obtain an "electron crystal") while hampering as much as possible the motion of phonons (a ``phonon glass'') \cite{Takabakate2014}. In other words, obstacles should be put  {in} the way of phonons that impede as little as possible the passage of electrons. 
}

{Within a polycrystal, obstacles to phonon motion include point defects---vacancies, interstitials, impurities---line defects such as dislocations, surface defects such as grain boundaries, interfaces and free surfaces, and what could be termed three dimensional defects such as precipitates. Recently, the role of dislocations has become the focus of much attention. For example, Shuai et al. \cite{Shuai2016} have reported high thermoelectric performance of Bi-based Zintl phases (Eu$_{0.5}$Yb$_{0.5}$)$_{1{-}x}$Ca$_x$Mg$_2$Bi$_2$, and have correlated this performance with an increase in the dislocation density of the material. In a similar vein, Wu et al. \cite{Wu2019} have added small amounts of Na, Eu and Sn to PbTe to obtain Na$_y$Eu$_{0.03}$Sn$_{0.02}$Pb$_{0.95 {-} y}$ Te. They have shown that, when $y= 0.03$, there is a significant increase of dislocation density, together with a significant decrease in thermal conductivity. You et al. \cite{You2018} have studied the behavior of  a PbSe-Cu system in which the presence of dislocations leads to a significantly decreased lattice thermal conductivity. Xin et al. \cite{Xin2017} have studied the thermal behavior of Mg$_2$Si$_{1{-}x}$Sb$_x$ and have concluded that an increase in dislocation density leads to a decrease in thermal conductivity. This strategy of introducing an extra alloying element into a given thermoelectric material has been followed by a number of researchers: Zhou et al. \cite{Zhou2018} have introduced Sb and Te into PbSe, while Yu et al. \cite{Yu2018} introduced Ag into PbTe. In both cases there was a decrease in the thermal conductivity that could be related to the presence of dislocations. {A numerical experiment, using molecular dynamics, has been carried out by Giaremis et al. \cite{Giaremis2020} in order to investigate the effect of decorated dislocations on the thermal conductivity of GaN. They have concluded that decorated dislocation engineering can lead to interesting fabrication strategies for themoelectric devices. } 
}

{The theory tool used by the researches mentioned in the previous paragraph is the classical analysis of Klemens~\cite{Klemens1955}. In this work, dislocations are considered as static, straight line, defects of infinite length. That is, they are point defects in a two-dimensional lattice, that  {extend themselves} into the third dimension to plus and minus infinity by  {a} homogeneous translation. Now, dislocations in any material have a finite length, a typical magnitude (except in especially designed materials, see below) being $\sim$ 100 nm or less. Also, dislocations are by no means static. They respond to an incoming elastic wave, i.e. to a phonon, by bowing out, and this response has been well-known and widely documented over decades~\cite{Granato1956a,Granato1956b,Anderson2017}.  {Admittedly,} there is no denying that the data mentioned in the previous paragraph can be fit with a contribution to the phonon inverse relaxation time that is a linear combination of  {terms} linear in frequency and cubic in frequency, as explained by Klemens. However, there does not appear to be a clear relation between the parameters needed to obtain a fit to the data and the parameters characterizing each specific material. From a different perpective, Wang et al. have performed ab-initio numerical calculations of the scattering of phonons by dipole dislocations in GaN~\cite{Wang2019} and in Si~\cite{Wang2017} concluding that, while said scattering is significant, it is not quantitatively accounted for by the model of Klemens~\cite{Klemens1955}. Clearly, a better modeling of the phonon-dislocation interaction is needed. In this paper we address this concern.
} 

{A rather significant development has been reported by Sun et al.~\cite{Sun2018}, who have fabricated a micron thick InN single crystal with a highly oriented dislocation array that  {pierces} through the film  {across}, and have measured the thermal conductivity both parallel and perpendicular to the film thickness. There is a factor as high as ten between the two, which is far larger than what would be expected from the single crystal anisotropy. Of course, free boundaries and point defects are the same for both directions. This anisotropy has been measured as a function of temperature and of dislocation density. The model of Klemens \cite{Klemens1955} can reproduce, roughly, the temperature dependence of the cross-plane thermal conductivity, but cannot reproduce its dependence on dislocation density;  {its prediction is} also quite far away from the  {observed} values for the in-plane thermal conductivity. 
}

{A modification of the model of Klemens \cite{Klemens1955} was worked out by Carruthers \cite{Carruthers1959} in order to provide a theoretical basis for a larger phonon-dislocation scattering cross section. While the theory of Carruthers \cite{Carruthers1959} provides a crude estimate of the results of Sun et al. \cite{Sun2018}, it does not appear to  {accurately} capture the temperature dependence, or the dislocation dependence,  {of the complete thermal conductivity tensor (involving both in- and cross-plane conductivities)}. In any case, the calculation of Carruthers \cite{Carruthers1959} is based on an anharmonic interaction between atoms that does not appear to have received independent validation.
}

{An additional feature of the experiment of Sun et al.~\cite{Sun2018} is that most heat-carrying phonons appear to} have wavelengths shorter than 5 nm, which is much smaller than the apparent typical length of a dislocation in their setup, $L \sim 1 \, \mu$m. Therefore, understanding the interaction between phonons and dislocations where the wavelength of the former is much smaller than the spatial extent of the latter should prove to be essential in describing and explaining thermal transport anisotropies mediated by dislocations. {In a previous publication~\cite{Lund2019} we have developed a quantum theory of phonons in interaction with dislocation segments of finite length. Here we extend that formalism to infinitely long dislocations, and explore the consequences for the thermal transport properties of a material threaded with many parallel such dislocations.}

{The dynamic response of an infinitely long dislocation to an incoming phonon was considered by Ninomiya~\cite{Ninomiya1968,Ninomiya1969}, who introduced a phonon-dislocation coupling through the kinetic energy terms in the system Hamiltonian, while not retaining a potential energy coupling. This interaction has been recently quantized by Li et al.~\cite{Li2017a,Li2017b,Li2018}. In this paper we shall quantize the phonon-dislocation interaction when the dislocation is an infinitely long elastic string, but the coupling, as explained in more detail in the next section, is through the potential energy term and is chosen to reproduce well-established classical results, as encapsulated by the Peach Koehler force~\cite{PeachKoehler1950,Anderson2017}.
}

{The possibility that oscillating---as opposed to static---dislocations should contribute to  thermal transport was considered by Granato~\cite{Granato1958} soon after the work of Klemens~\cite{Klemens1955}. However, no satisfactory agreement with experimental measurements could be found. In retrospect, it would appear that one difficulty with the theory, as gleaned from the 1982 review by Kneezel and Granato~\cite{Kneezel1982}, was that the phonon scattering rate induced by the moving dislocations was calculated to be proportional to a damping term in the dislocation dynamics, itself proportional to dislocation velocity, with an uncontrolled proportionality constant. In the present paper, we introduce no such phenomenological parameters. On the contrary, our theoretical framework sets the stage for the calculation of velocity-dependent damping of long wavelength phonons by their interaction with short wavelength dislocation oscillations. We shall briefly touch on this issue below.
}

{This paper is organized as follows: Section \ref{Sec:classical_action} recalls the classical theory that will be quantized. Section \ref{canonical_quant} introduces the canonical quantization of the free phonons and free dislocation modes which, following Li et al.~\cite{Li2017a} we shall call ``dislons''. Section \ref{sec:quad} introduces the interaction between phonons and dislons and presents a number of consequences of this interaction, calculated to lowest order (classically, this means for small strains). The basic tools are described, and the $\T$ matrix for the scattering of a phonon by a dislocation is calculated  in Section \ref{sec:fd}, while Section \ref{sec:disl-prop}   discusses how the particle-like properties of dislons are modified by their interaction with phonons. An interesting result is the computation of the phonon contribution to the phenomenological damping (the so-called ``$B$'' term) that is introduced in classical descriptions of dislocation  {dynamics} as elastic strings~\cite{Granato1956a,Maurel2005a}.  {The} consequences of  {the aforementioned interaction} for thermal transport are worked out in Section \ref{sec:thtr}, with special attention to the case of a solid threaded by a large number of parallel dislocations. Phonon scattering cross sections and lifetimes are discussed in sections~\ref{sec:csecs} and~\ref{sec:ltimes}, respectively. Section~\ref{sec:KC} compares the consequences with the classical results of Klemens and Carruthers.  {One} striking difference is that, while in the Klemens/Carruthers approach the phonon lifetime is  {proportional} to frequency and, if the contribution of the dislocation core is considered, to frequency cubed, our approach leads to a phonon lifetime that is {\it inversely proportional} to frequency, as discussed  {in} Section \ref{sec:KC}. A quantitative comparison of the resulting anisotropy in thermal transport is given in Section \ref{sec:anisotropy}. Finally, Section \ref{sec:conclusions} has a discussion, outlook and concluding remarks.
}

\section{Classical action}
\label{Sec:classical_action}
In this work we work out the quantum theory of oscillating dislocation segments, of infinite length, in interaction with elastic waves in three dimensions. {The quantum theory to  be constructed is based on a well-established classical theory, whose principal aspects we recap here.}

We consider a homogeneous, solid, elastic continuum of density $\rho$, possibly anisotropic along one axis $\hat{e}_3$, and elastic constants $c_{pqmr} \, (p,q = 1,2,3)$. Within the solid there is a string-like dislocation line. The variables describing the solid are the displacements  ${\mathbf u (\mathbf x,t)}$, at time $t$, of a point whose equilibrium position is ${\mathbf x}$.  The string is  described by a vector ${\mathbf X} (s,t)$, where $-\infty <s< \infty $ is the coordinate along the string equilibrium axis, whose  {endpoints} are fixed at spatial infinity. {The motion of this string is  {one} of small amplitude away from an equilibrium position that is a static straight line  {parametrized by} ${\mathbf X}_0 (s) = s\, \hat e_3 $.} The fact that the string is a dislocation is implemented by the displacements ${\mathbf u (\mathbf x,t)}$ being multivalued functions: they have a discontinuity equal to the Burgers vector $\b$ when crossing a surface whose boundary is the string. In addition to this geometrical fact, the coupling between elastic displacements and elastic string is given by standard conservation of energy and momentum arguments \cite{Lund1988}. When dislocation velocities are small compared to the speed of sound, an assumption we shall make throughout this work, this leads to the well-known Peach-Koehler force \cite{PeachKoehler1950}.  
In the time-dependent case, and for string velocities small compared to the speed of sound, the dynamics is described by the following classical action: 

\beq
S= S_{\rm ph} + S_{\rm string} + S_{\rm int} + S_0
\label{actionFL}
\eeq
where 
\bea
S_{\rm ph} &=& \frac{1}{2} \int \! dt \! \int \! d^3 x  \! \left( \rho \dot{\mathbf u}^2 - c_{pqmr} \frac{\partial u_m}{\partial x_q} \frac{\partial u_p}{\partial x_r} \right) + \cdots  \label{action1}\\
S_{\rm string} &=& \frac{1}{2} \int \! dt \! \int_{-\infty}^{\infty} \!\!\! ds  \! \left( m \dot{\mathbf X}^2 - \Gamma {\mathbf X}'^2 \right) + \cdots \label{action2}\\
S_{\rm int} &=& - b_i \int \! dt \! \int_{\delta \mathcal{S}} dS^j \sigma_{ij}. \label{action3}
\eea
where the ellipses ``$\cdots $'' refer to higher order terms in the phonon or string actions  {in the sense that they involve higher powers of the dynamic fields ${\mathbf u}$ and ${\mathbf X}$. Also,} $b_i$ is the $i$-th component of the Burgers vector, $\sigma_{ij}$ is the elastic stress tensor, {evaluated at the current position of the dislocation line,} and the surface $\delta \S$ describes the region bounded by the string and its equilibrium position.  {Finally, } $S_0$ involves the interaction of the elastic displacements with a static dislocation, whose boundary is the straight line ${\mathbf X}_0$. It does not contribute to the dynamics and shall be ignored in the sequel.

Let us start by describing the phonon action $S_{\rm ph}$. Clearly, $S_{\rm ph}$ describes elastic waves in an elastic continuum, wherein the quadratic terms will lead to free phonons and the higher order terms will lead to phonon-phonon interactions. In the isotropic case, $c_{pqmr} = \lambda \delta_{pq} \delta_{mr} + \mu (\delta_{pm} \delta_{qr} + \delta_{pr} \delta_{mq} )$ where $\lambda$ and $\mu$ are the Lam\'e constants.  {However,} in the axially anisotropic case, for a medium with hexagonal or transverse isotropic symmetry, there are five independent elastic constants.

In the absence of the interaction term, the free phonon theory has simple solutions in terms of plane waves for the elastic displacement (phonons) and normal modes for the elastic  {string-like dislocation}. In the case of phonons, they are in general characterized by three different modes of propagation,  {determined by the wave equation
\beq
\rho \ddot{u}_p = c_{pqmr} \frac{\partial^2 u_m}{ \partial x_r \partial x_q}.
\eeq}

In the completely isotropic case, we have only two  {distinct} modes: transversal waves of speed $c_T = \sqrt{\mu/\rho}$ with two allowed polarizations, and longitudinal waves of speed $c_L = \sqrt{(\lambda+2\mu)/\rho}$ with one polarization. 

On the other hand, we have a string described by a vector  {field} ${\mathbf X} (s,t)$, where $-\infty<s<\infty$ is a position parameter along  {said} string. We consider small deviations from a straight equilibrium position ${\mathbf X}_0$, the ends of which are pinned to their positions at infinity. $S_{\rm string}$ describes oscillations (normal modes) of an elastic string of infinite length with fixed ends; higher order terms describe anharmonic effects on these oscillations, which we will not address in this work. The parameters $m$ (mass per unit length) and $\Gamma$ (line tension) characterize the dislocation segment.

In previous  {work~\cite{Lund2019}}, considering only the completely isotropic case, we have assumed segments of edge dislocations only, in which case they may be written in terms of the Burgers vector $\b$ and $\gamma \equiv c_L/c_T$ as 
\beq
m_{\rm isotropic} = \frac{\rho b^2}{4\pi} (1 + \gamma^{-4}) \ln (\delta/\delta_0)
\label{eq:dislocmass}
\eeq
where $\delta$, $\delta_0$ are long- and short-distance cutoff lengths, and
\beq
\Gamma_{\rm isotropic} = \frac{\mu b^2}{2\pi} (1 - \gamma^{-2}) \ln (\delta/\delta_0).
\label{eq:disloctension}
\eeq
 {For an anisotropic solid, one can expect that the factors $(1 + \gamma^{-4})$ and $(1 - \gamma^{-2})$ in $m$ and $\Gamma$ will be modified to reflect the specific geometry of the solid, but should otherwise remain unchanged.}

Classically, phonons have plane wave solutions. Similarly, the string term $S_{\rm string}$ of the action leads to oscillatory solutions that may be expanded in Fourier series ${\rm Re}\left\{ \int \frac{d\kappa}{2\pi} a(\kappa) e^{- i\omega_\kappa t} e^{-i \kappa s } \right\}$, where $\omega_\kappa = \kappa \sqrt{\frac{\Gamma}{m}}$ is the frequency of each normal mode. We take the string to have one degree of freedom (i.e. one direction orthogonal to its equilibrium position over which to oscillate) defined by the direction of the Burgers vector $\b$,  {thus defining} the glide plane. For most of our results, the generalization to more directions of oscillation is straightforward.

Finally, $S_{\rm int}$ describes the interaction between these two sectors. It is straightforward to check that
\beq
\frac{\delta S_{\rm int}}{\delta X_k} = {-b_i \e_{jkm} X'_m \sigma_{ij}}
\eeq
 {reproduces} the well known Peach Koehler force~\cite{PeachKoehler1950,Anderson2017}. {This phonon-dislocation coupling has been successfully used for decades~\cite{Granato1956a,Granato1956b}. In recent years, it has been used to compute the scattering cross section of elastic waves by dislocation segments in a first Born approximation, a result that has been further employed to compute the change in propagation velocity and attenuation for said waves by many such dislocation segments~\cite{Maurel2005b,Natalia2009,Churochkin2016}. These results, in turn, have led to novel ways of acoustically characterizing the plasticity of metals and alloys~ \cite{Mujica2012,Barra2015,Salinas2017,Espinoza2018}}.

Having  {reviewed} the classical theory, we now turn to its quantization by introducing canonical commutation relations.  {We note that} the theoretical setup just described includes anisotropic media. Even though we shall work out specific quantitative consequences only in the isotropic case, we shall keep the notation, as far as possible, compatible with anisotropy. 

\vspace{5mm}

\section{Canonical quantization of the free fields}   
\label{canonical_quant}

We commence this section by implementing the correspondence of Poisson brackets to commutators $\{ \cdot , \cdot \} \to -\frac{i}{\hbar} [ \cdot , \cdot]$. According to standard practice~\cite{Srednicki2007,Peskin1995},
the mode coefficients of the classical solutions are promoted to creation and annihilation operators, in terms of which we may write the displacement field as
\bea
\u (\x,t) &=& \sqrt{\frac{\hbar}{\rho}} \! \int \! \frac{d^3k}{(2\pi)^3} \!  \sum_{\iota \in \{{\rm pol.}\}}  \left[ \frac{\e_{\iota}^*(\k) a_{\iota}(\k) e^{ i\k \cdot \x -i\omega_{\iota}\! (\k)t}}{\sqrt{2 \omega_{\iota}\! (\k) }} \right. \nonumber \\  
&& \hspace{6em} \left.
 + \frac{\e_{\iota}(\k) a_{\iota}^{\dagger}(\k) e^{ - i\k \cdot \x +i\omega_{\iota}\! (\k)t}}{\sqrt{2 \omega_{\iota}\! (\k) }}   \right],
 \label{qdispl}
\eea
 {where the sum over $\iota$ represents the sum over phonon polarizations. On the other hand, we may write} the string displacement as
\bea
X(s,t) &=& \sqrt{\frac{\hbar}{m}} \int \frac{d\kappa}{2\pi} \left( \frac{\alpha(\kappa) e^{-i\omega_\kappa t} e^{-i \kappa s }}{\sqrt{ 2 \omega_\kappa }} \right. \nonumber \\   && \hspace{5em} \left.
 +  \frac{\alpha^{\dagger}(\kappa) e^{i\omega_\kappa t} e^{i \kappa s }}{\sqrt{2 \omega_\kappa}}   \right).
\eea

Now we proceed to impose canonical commutation relations:
\bea
[u_i(\x,t), \rho {\dot u}_j(\y,t)] &=& i \hbar \delta_{ij} \delta^{(3)}(\x - \y), \\   
{[}X(s,t), m {\dot X}(s',t){]}  &=& i \hbar \delta(s-s'), \\
{[}u_i(\x,t), u_j(\y,t){]} = 0, & & {[}X(s,t), X(s',t){]} = 0,
\eea
which fully define the quantum theory in the non-interacting case. These relations in turn require
\begin{align}
[a_{\iota}(\k), a_{\iota'}(\k')] &= [a_{\iota}^{\dagger}(\k), a_{\iota'}^{\dagger}(\k')] = 0, \\
[a_{\iota}(\k), a_{\iota'}^{\dagger}(\k)] &= (2\pi)^3 \delta^{(3)}(\k - \k') \delta_{\iota \iota'}, \\
[\alpha(\kappa), \alpha(\kappa')] &= [\alpha^{\dagger}(\kappa), \alpha^{\dagger}(\kappa')] = 0, \\
[\alpha(\kappa),  \alpha^{\dagger}(\kappa')] &= (2\pi) \delta^{(1)}(\kappa - \kappa').
\end{align}
In the preceding expressions, $\iota$ is an index that runs over the possible polarizations for the phonons, which  {in the isotropic case goes over two transverse polarizations that we will denote by $\iota = T_1, T_2$, and one longitudinal polarization that we will denote by $\iota = L$. In the anisotropic case there would be three inequivalent polarizations: transverse polarization with displacements within the $\hat e_1 - \hat e_2$ plane, transverse polarization within the $\hat e_3 - \hat k$ plane, and longitudinal polarization inside the $\hat e_3 - \hat k$ plane. } $\e_\iota(\k)$ represents the polarization vector associated to each mode of propagation.
The corresponding  {eigenfrequencies} satisfy $\omega_{\iota}(\k) = c_\iota( \hat k \cdot \hat e_3 ) k$, with  {two phase velocities $c_T, c_L$ in the isotropic case, and three phase velocities} that depend on the angle between the wave-vector and the anisotropy axis  {in the anisotropic case. Finally,} $\omega_\kappa = \kappa \sqrt{\Gamma/m}$ is the frequency for the mode of the string with wavenumber $\kappa$.

{To complete the description of the theory, we need to specify its dynamics, which are generated by the time-evolution implied by a Hamiltonian operator. In the case of the ``free'' theory, where no interactions between phonons and ``dislons'' (the excitations on the string) take place, the Hamiltonian is}
\beq
H = H_{\rm ph} + H_{\rm string}
\eeq
with phonon and string terms given, respectively, by
\bea
 H_{\rm ph} & = & \int \! \frac{d^3k}{(2\pi)^3} \!  \sum_{\iota \in \{{\rm pol.}\}} \hbar \omega_{\iota}(\k) a_{\iota}^{\dagger}(\k) a_{\iota}(\k),\\
 H_{\rm string} & = &  \int \frac{d\kappa}{2\pi} \hbar \omega_\kappa  \alpha^{\dagger}(\kappa)\alpha(\kappa).
 \eea

In characterizing the free theory, a fundamental object is the two-point function, more commonly known as the \textit{propagator}. Let $\text{T}$ be the time-ordering symbol, instructing operators evaluated at a later time to be placed at the left, and let $\ket{0}$ be the vacuum state of the quantum mechanical system, with no excitations of the elastic displacements nor the string. For the dislocation, it reads
\begin{equation}
\begin{split}
\Delta(s-s',t-t') \equiv & \braket{0 | \text{T} X(s,t) X(s',t') | 0} \\ = & \frac{\hbar}{m}  \int \frac{d\kappa}{2\pi} \frac{e^{-i \omega_\kappa |t-t'| }}{2 \omega_\kappa} e^{i \kappa (s - s')}.
\end{split}
\end{equation}
Even though we can do the same for the elastic displacement field, it turns out to be more useful for subsequent computations to write down the propagator for its spatial derivative:
\begin{equation}
\begin{split}
\Delta_{iji'j'} & (\x - \x', t- t') \equiv  \braket{0 | \text{T} \frac{\partial u_i}{\partial x_j}(\x,t) \frac{\partial u_{i'}}{\partial x_{j'}}(\x',t') | 0} \\ = & \frac{\hbar}{\rho} \int \frac{d^3k}{(2\pi)^3} \sum_{\iota \in \{{\rm pol.}\}} \\ & \quad \times  k_j k_{j'}    \e_\iota(\hat k)_i \e_\iota^*(\hat k)_{i'}  \frac{e^{-i \omega_\iota(\k) |t - t'|} }{2 \omega_\iota(\k)} e^{i \k \cdot (\x - \x')}
 \\ = & \frac{\hbar}{\rho} \int \frac{d^3k}{(2\pi)^3} \int \frac{d\omega}{2\pi i} \sum_{\iota \in \{{\rm pol.}\}}  \\ & \quad \times  \frac{k_j k_{j'} \e_\iota(\hat k)_i \e_\iota^*(\hat k)_{i'}}{- \omega^2 + \omega_\iota(\k)^2 - i \epsilon }  e^{i \k \cdot (\x - \x')} e^{-i \omega (t - t')},
\end{split}
\end{equation}
where $\epsilon$ is a positive infinitesimal.

The reason behind writing down time-ordered quantities is that when we compute scattering amplitudes in the interacting theory, we  {will be} interested in the $S$-matrix, given by \cite{Srednicki2007,Peskin1995}
\beq \label{amplitude-basic}
\braket{\Psi_{\rm out} | \text{T} \exp \left[ - \frac{i}{\hbar} \int_{-\infty}^{\infty} H_I(t) dt \right] \! |\Psi_{\rm in} }
\eeq
where $H_I$ is the quantum mechanical interaction picture Hamiltonian operator.  {Therefore, if we expand the exponential in~\eqref{amplitude-basic} in a power series of $H_I$, all terms in the series will be time-ordered; thus giving the computation of} time-ordered quantities a central role.

{In what follows, we will denote by $u^{I}$ and $X^{I}$ the operator fields associated to lattice displacements and to the oscillations of the string-like dislocation,  {respectively,} in the interaction picture of quantum mechanics, which evolve as free fields. As a reminder to the reader, the passage  {between the interaction picture and} the Heisenberg picture is implemented through
\beq
u (\x , t) = U^\dag (t,t_0)  u_I (\x , t)  U (t,t_0)
\eeq
with
\beq
U (t,t_0) = \text{T} \exp \left[ - i \int_{t_0}^{t} H_I (t') d t' \right]
\eeq
where $t_0$ is the time at which both operators coincide (typically  {in a scattering context} it is taken to be $-\infty$).}

 {Having set up the formalism, we can now dive into the interaction Hamiltonian $H_I$ of interest and explore the dynamics it generates for the constituents of our theory: phonons and dislons.}

\section{The quadratic interactions with a single string} \label{sec:quad}

The lowest order interaction  {out of $S_{\rm int}$~\eqref{action3}}, which classically means to consider small strains and small string excursions away from the equilibrium position, is given by
\beq
S_{\rm int}^{(2)} = -N b \int dt   \int_{-\infty}^\infty \!\!\!\! ds \, {\bf M}_{kl} \frac{\partial u_k}{\partial x_l}(\x_0+(0,0,s),t) X(s,t) 
\eeq
which is quadratic in the fluctuations.  {In this expression, we have defined $N \equiv c_{1212}$. We will take the string to have its equilibrium position along the $ \hat{e}_3$ axis, and the burgers vector to be written as $\b = b \hat{e}_1$.} With  {this} choice of coordinates, ${\bf M}_{kl}= (\hat{e}_1)_k (\hat{e}_2)_l +  (\hat{e}_2)_k (\hat{e}_1)_l $.  This interaction will give rise to the scattering of phonons by the string, which is described by
\beq \label{ph-ph-a}
\langle f|i\rangle = \braket{0| a_{\iota'}(\k') \text{T} \exp \left[ - \frac{i}{\hbar} \int_{-\infty}^{\infty} H_I(t) dt \right] a_{\iota}^{\dagger}(\k)  |0},
\eeq
where
\beq \label{quad-hamiltonian}
H_I(t) = N b   \int_{-\infty}^\infty \!\!\!\! ds \, {\bf M}_{kl} \frac{\partial u^{I}_k}{\partial x_l}(\x_0+(0,0,s),t) X^{I}(s,t),
\eeq
 {and $a_{\iota}^\dagger(\k)$, $a_{\iota'}(\k')$ are creation and annihilation operators that define the initial (one phonon with wavenumber $\k$ and polarization $\iota$) and final (one phonon with wavenumber $\k'$ and polarization $\iota'$) states.}

For completeness, we also write down the Hamiltonian in the Heisenberg picture (where it is naturally constant) in terms of creation and annihilation operators:
\beq
\begin{split}
H_{\rm int} = \, & \hbar \int \frac{d^3k}{(2\pi)^3}  \sum_{\iota \in \{ {\rm pol.} \} } \\ & \times \left(\frac{i E(\k;\iota) a_\iota(\k)}{\sqrt{2 \omega_{k_3} } } ( \alpha(-k_3) + \alpha^{\dagger}(k_3) ) + {\rm h. c.} \right)
\end{split}
\eeq
where h.c. stands for hermitian conjugate. {Note that this is a quadratic interaction.} Furthermore, following previous work~\cite{Lund2019}, we have defined
\beq
E(\k;\iota) \equiv \frac{N b}{\sqrt{2\rho m \omega_{\iota}(\k) }} k_l {\bf M}_{kl} \varepsilon_\iota^*(\k)_k   e^{i \x_0 \cdot \k},
\eeq
and we will denote its complex conjugate by $E^*$. Here we have denoted $k_l = (\k \cdot \hat{e}_{l})$ and $\varepsilon_\iota(\k)_k = (\e_\iota(\k) \cdot \hat{e}_k)$.

We now turn to the main question of interest in this article: how does a (comparatively) short wavelength phonon with wavenumber $\k$ and polarization $\iota$ propagating through the elastic continuum interact with a long (approximately infinite) dislocation segment with length $L$?  {To answer this question, we proceed as follows: In section~\ref{sec:fd} we organize and solve the theory in terms of Feynman diagrams, to then study the properties of dislons as particles and scatterers in section~\ref{sec:disl-prop}. Finally, we briefly discuss how the phonon-dislon interaction makes both excitations reach thermal equilibrium in section~\ref{sec:th-equil}, before moving on to the implications of this interaction on thermal transport in the subsequent section.}

\subsection{Phonon by dislocation scattering: amplitudes and Feynman diagrams}
\label{sec:fd}

In this section we describe how to obtain the scattering amplitude of a phonon by a dislocation, to all orders in the interaction Hamiltonian~\eqref{quad-hamiltonian}. That is, we explicitly perform the computation of all terms in the power series development of the exponential in (\ref{ph-ph-a}). Since we have a quantum field theory in our hands, it is natural to carry out the computation in terms of Feynman diagrams. This is a powerful method to organize the various terms that appear in scattering processes.

\begin{figure}[t]
\includegraphics[width=0.5\textwidth]{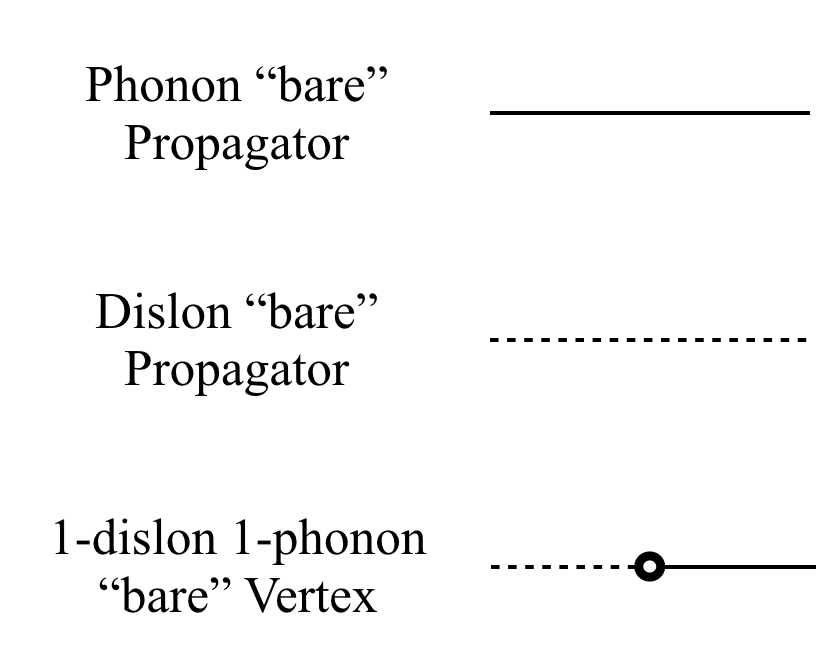} 
\caption{The basic diagrammatic expressions to be used in solving the quadratic theory. The upper two diagrams describe the basic propagator (Green's function) for each type of excitation in the solid, with a continuous line for phonons and a  {dashed} line for dislons. The third diagram represents the quadratic phonon-dislon interaction, allowing for quantum of excitations of one type to be converted into the other.}  \label{fig:diagrams-quad}
\end{figure}

In the quadratic theory, the basic diagrammatic elements are those shown in Figure~\ref{fig:diagrams-quad}, each representing a specific contribution that gives form to the  {scattering processes}. They are:
\begin{itemize}
\item Phonon ``bare'' propagator: the phonon Green's function. In particular, we will be more interested in its derivatives, or the earlier defined $\Delta_{iji'j'} (\x - \x', t- t')$ whose value in momentum-frequency space is given by
\beq
\Delta_{iji'j'}(\k,\omega) = \sum_\iota \frac{-i \hbar k_j k_{j'} \varepsilon_\iota(\hat k)_i \varepsilon_\iota^*(\hat k)_{i'} }{- \rho \omega^2 + \rho c_\iota^2(\hat k) k^2 - i\epsilon },
\eeq
where we have omitted an overall Dirac delta imposing momentum-frequency conservation $(2\pi)^4 \delta(\omega - \omega') \delta^3(\k - \k')$.
\item Dislon ``bare'' propagator: the dislon Green's function. It is given in momentum-frequency space by
\beq
\Delta(\kappa, \omega) = \frac{-i \hbar}{ -m \omega^2 + \Gamma \kappa^2  - i\epsilon },
\eeq
where we have also omitted an overall Dirac delta imposing momentum-frequency conservation $(2\pi)^2 \delta(\omega - \omega') \delta(\kappa - \kappa') $.
\item 1-dislon 1-phonon ``bare'' vertex: diagrammatical representation of the quadratic interaction between phonons and dislons. In position space, it instructs to operate over the coincident position coordinate of both adjacent propagators as
{\beq
\begin{split}
i {\bf V}_2 &= \frac{-i b c_{1212} {\bf M}_{i'j'}}{\hbar} \\ & \times \int_{-\infty}^\infty \!\!\!\! ds \Delta_{ \textcolor{gray}{ij} i'j'}\textcolor{gray}{(\x -} (0,0,s),\textcolor{gray}{ t -} t')  \Delta(s \textcolor{gray}{- s'}, t' \textcolor{gray}{- t'')  }.
\end{split}
\eeq}
This coupling will have important consequences when we examine the exact propagator for the dislocation excitations. Moreover, it enforces momentum (wavenumber) conservation along the direction of the dislocation line, through a factor  $(2\pi) \delta(\kappa - k_3')$ that appears after integrating over the string coordinate $s$.
\end{itemize}

\begin{figure*}[t] 
\includegraphics[width=1\textwidth]{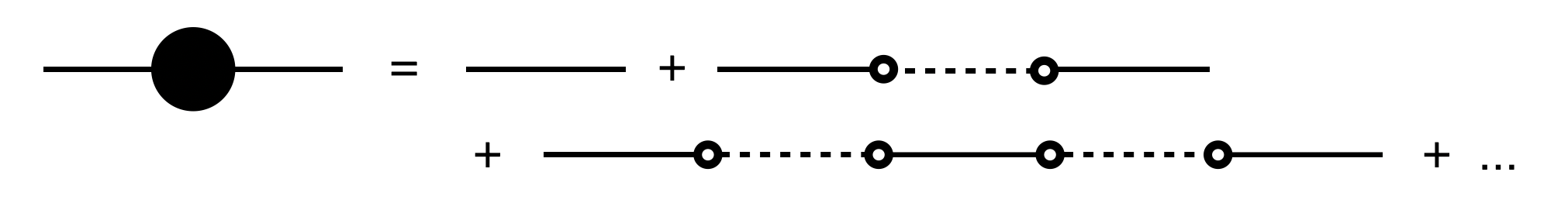}
\caption{ ``Dressed'' phonon propagator. This is the main composite object that appears in the quadratic theory calculations, and it does so as a consequence of incorporating the consequences of  {the} interaction to all orders in perturbation theory.}
\label{fig:propagator-a}
\end{figure*}

With these tools, we want to evaluate~\eqref{ph-ph-a}, which corresponds to having one ingoing phonon and one outgoing phonon as external states, which in the diagrams  {that represent our scattering process are depicted by the ``external'' lines (i.e., those that have one of their ends not attached to another diagrammatic piece)}. In essence, we want to compute a ``dressed'' propagator for the elastic displacement field, which determines the probability of measuring a phonon with wave-vector $\k'$ as a result of having sent in a phonon with momentum $\k$ into the elastic medium. This is schematically represented in Figure~\ref{fig:propagator-a}.

{The computation can now be organized by the number of vertices (i.e., insertions of the interaction-picture Hamiltonian as a result of expanding the time evolution operator in a power series) present in each diagrammatic piece of Figure~\ref{fig:propagator-a}. By the means of Wick's theorem~\cite{Srednicki2007,Peskin1995}, one can then evaluate the expectation value corresponding to the transition amplitude~\eqref{ph-ph-a}, which essentially contracts all interaction-picture fields in pairs in all possible ways amongst themselves.}

\begin{figure}[b] 
\includegraphics[width=0.4\textwidth]{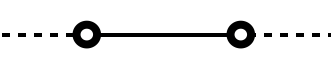}
\caption{ Diagrammatic representation of the dislon self-energy $\Pi$, where the dashed ``external'' lines carry ingoing and outgoing momenta $\kappa$ and frequency $\omega$.}
\label{fig:dislon-self}
\end{figure}

{This contraction in all possible ways effectively gives an $n!$ factor that cancels the $1/n!$ in the Taylor expansion of the exponential, because the time-ordering symbol makes all contractions equivalent at a fixed order in the perturbative series. Furthermore, phonon propagators conserve 3-momentum (the wavenumber $\k$) and frequency/energy $\omega$, whereas dislon propagators only conserve momentum along the dislocation axis $k_3 \equiv \k \cdot \hat e_3$ and energy $\omega$. Then, since the integrals in the interaction-picture time evolution operator (one over time and one over the longitudinal extent of the dislocation $s$) propagate the conservation of $k_3$ and $\omega$, all one needs to do is work out the pieces in the diagram in Figure~\ref{fig:propagator-a} where modes with $k_1$ and $k_2$ components appear as ``virtual" intermediate states. This essentially amounts to calculating the \textit{self-energy} $\Pi(\kappa,\omega)$ of the dislon propagator, as is depicted in Figure~\ref{fig:dislon-self}.  {After evaluating this quantity,} the remainder of the computation will be given by summing over the number of intermediate dislon propagators {to construct the \textit{exact} dislon propagator,
\beq
\begin{split}
S(\kappa, \omega) &\equiv \Delta(\kappa,\omega) \sum_{n=1}^\infty \left( \frac{i}{\hbar} \Pi(\kappa,\omega) \Delta(\kappa,\omega) \right)^{n-1} \\ 
&= \frac{1}{\Delta^{-1}(\kappa,\omega) - \frac{i}{\hbar} \Pi(\kappa,\omega) },
\end{split}
\eeq
which includes the effect of all the intermediate phonon states in a scattering process.}
}

{Using the diagrammatic elements we defined earlier, and letting $(\kappa,\omega)$, $(\kappa',\omega')$ label the ingoing and outgoing states respectively, we obtain the following expression for the dislon self-energy
\beq
\begin{split}
& (2\pi) \delta(\omega - \omega') (2\pi) \delta(\kappa - \kappa') \frac{i}{\hbar} \Pi(\kappa,\omega) \\  &  = \frac{(-i)^2 N^2 b^2}{\hbar^2} {\bf M}_{kl} {\bf M}_{k'l'} \!\! \int_{-\infty}^\infty \!\!\!\!\! dt \int_{-\infty}^\infty \!\!\!\!\!  dt' \int_{-\infty}^\infty \!\!\!\!\!  ds \int_{-\infty}^\infty \!\!\!\!\!  ds'  \\ & \quad \times   \int \!\! \frac{d^3 k}{(2\pi)^3} \!\! \int \frac{d\omega''}{2\pi i} \Delta_{klk'l'}(\k,\omega) e^{i (s - s') k_3} e^{-i \omega'' (t - t')}  \\ & \quad \quad \quad \quad \quad \quad \quad \quad \quad \quad \quad \quad \,\,\, \times e^{i\omega t -i \omega' t'} e^{-i \kappa s + i \kappa' s'} .
\end{split} 
\eeq
}

{The integrals over time and $\omega''$ are straightforward and give energy conservation. Some algebra allows one to compute the integral over the azimuthal angle and obtain
\beq
\begin{split}
& (2\pi) \delta(\kappa - \kappa') \Pi(\kappa,\omega)  \\ & \quad =  \frac{ N^2 b^2 }{16 \pi^2 \rho} \int_{-\infty}^\infty \!\!\!\!\!  ds \int_{-\infty}^\infty \!\!\!\!\!  ds' \int_{-\infty}^\infty \!\!\!\!\! dk \int_{-1}^1 \!\!\! du e^{i k u (s - s') } \\
& \quad \quad  \times e^{-i \kappa s + i \kappa' s'} k^4 (1 - u^2) \sum_\iota \frac{ |\varepsilon_\iota^{xy}(u) |^2 }{ c_\iota^2(u) k^2 - \omega^2 - i\epsilon  },
\end{split}
\eeq
where $u \equiv \cos \theta$ and we have extended the radial integral over $k$ to $-\infty$ as the integrand is symmetric under $k \to -k$ and $u \to -u$.  {In this equation we have introduced the quantity $|\varepsilon_\iota^{xy}(u)|$, which stands for the magnitude of the projection of the polarization vector $\e_\iota$ into the $\hat e_1 - \hat e_2$ plane.}
}

{From here, one proceeds to evaluate the integral over $k$ by contour integration, closing on the upper-half plane if $u(s-s') > 0$, and on the lower-half plane otherwise. After a straightforward, but tedious calculation, one arrives at
\beq
\Pi(\kappa,\omega) = m \omega^2 F(\kappa/\omega),
\eeq
}
where the function $F$ is given by
\beq \label{F}
F(x) = \frac{N^2 b^2}{8 \pi \rho m} \sum_{\iota \in \{{\rm pol.}\}} \int_0^1 du \frac{|\varepsilon_\iota^{xy}(u) |^2 }{c_\iota^4(u)} \frac{2u (1 - u^2) }{  x^2 c_\iota^2(u) - u^2 - i\epsilon  }.
\eeq
This integral can be cast as the function  {in the integrand} evaluated at a given point plus a principal part, but since in evaluating Cauchy principal values it is typical to include a small parameter, we stick to the complex representation given by~(\ref{F}), which already gives both the imaginary and real parts of the result.

\begin{figure*}[t] 
\includegraphics[width=1\textwidth]{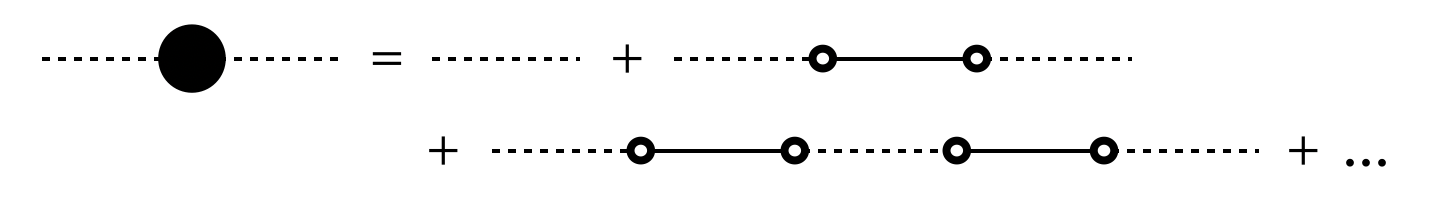}
\caption{ ``Dressed'' dislon propagator. This is the counterpart of the ``dressed'' phonon propagator in terms of the dislon field. This object has the advantage that it explicitly conserves energy and momentum along all dimensions in which the string is extended, whereas the exact phonon propagator only conserves  {energy and} momentum along the  {dimensions that are shared with the string (time and the $\hat e_3$ axis).}}
\label{fig:propagator-d}
\end{figure*}

For concreteness, we note that in the isotropic case this function can be calculated easily {in terms of relatively simple functions}. In terms of its real and imaginary parts, it would read:
\beq \label{Re-F-iso}
\begin{split}
{\rm Re} \left\{ F(x) \right\} &= \, \frac{ \mu^2 b^2}{8\pi \rho m c_T^4} \! \left[ \frac{1}{2} + \frac{3}{2 \gamma^4} + \left(1 - \frac{1}{\gamma^2} \right) c_T^2 x^2 \right. \\ & \quad \quad \quad \quad \left.   - (1 - c_T^4 x^4) \ln \! \left( \frac{|1 - c_T^2 x^2|}{|c_T^2 x^2|} \right)  \right. \\ & \quad \quad \quad \quad \left.  - \frac{(1 - \gamma^2 c_T^2 x^2)^2}{\gamma^4} \ln \! \left( \frac{|1 - \gamma^2 c_T^2 x^2|}{| \gamma^2 c_T^2 x^2|} \right)   \right]
\end{split}
\eeq
\beq \label{Im-F-iso}
\begin{split}
{\rm Im} \left\{ F(x) \right\}  = \frac{\mu^2 b^2}{8 \rho m c_T^4}  & \Big[ (1 - c_T^4 x^4) \Theta(1- |c_T x|) \\ &  + \frac{(1 - \gamma^2 c_T^2 x^2  )^2 }{\gamma^4} \Theta(1-|\gamma c_T x|)    \Big],
\end{split}
\eeq
where $\Theta(x)$ is the Heaviside step function.
Note that both the real and imaginary parts of $F$ are nontrivial. This means that we will have anomalous dispersion with contributions to both the effective  {speed of propagation} of the  {dislons} and  {to} their decay rate.

{Once we have the self-energy, we can sum over all possible insertions of phonon propagators (see Fig.~\ref{fig:propagator-d}) to write the exact dislon propagator as 
\beq \label{disl-prop-ex}
S(\kappa,\omega) = \frac{-i \hbar}{-m \omega^2 + \Gamma \kappa^2 - \Pi(\kappa,\omega) },
\eeq
which is, apart from the contractions with external phonons in the amplitude~\eqref{ph-ph-a}, all we needed to compute. With this in hand, we can now write the final result for the amplitude in a closed form}. Specifically, if we define the $\T$ matrix through
\beq
\braket{f|i} \equiv i (2\pi)^2 \, \delta(\omega_{\iota}(\k) - \omega_{\iota'}(\k')) \, \delta^{(1)}(k_3 - k_3') \, \mathcal{T}, 
\eeq
then we obtain
\beq \label{scatt}
\T =  \frac{E^*(\k;\iota) E(\k';\iota') }{-\left( 1 + F \! \left( \frac{ k_3}{\omega} \right)  \right) \omega^2 + \frac{\Gamma}{m} k_3^2  },
\eeq
{where we have omitted the free theory result,  {i.e., a pure phonon propagator,} as it does not represent a scattering process.}
{ {Having established the form of the phonon-to-phonon scattering amplitude}, we can now compute the scattering cross-section of phonons due to the presence of a dislocation line. However, before doing that we will explore some aspects of this result to build some intuition on the physics behind it.}

\subsection{{A look at the dislon dispersion relation}} \label{sec:disl-prop}

{The exact dislon propagator~\eqref{disl-prop-ex}, besides its relevance for computing phonon scattering amplitudes, also gives us the possibility to study the dispersion relation of the modes that propagate along the dislocation line directly. Indeed, the physical ``asymptotic'' states in a scattering picture, which are made of a superposition of ``on-shell'' states, are usually characterized by a dispersion relation $\omega(\kappa)$ determined by the (possibly complex) poles of the propagator. In the absence of a medium with which to interact, the dislon dispersion relation is defined by
\beq
-m\omega^2 + \Gamma \kappa^2 = 0 \implies \omega^2(\kappa) = \frac{\Gamma}{m} \kappa^2.
\eeq
This means, for example, that perturbations with frequency $\omega$ will propagate along the dislocation line with a wavenumber given by $ \kappa = \pm \sqrt{m \omega^2/\Gamma}$, corresponding to the usual picture of wave propagation on a string.}

{However, the presence of a non-trivial self-energy $\Pi(\kappa,\omega)$ complicates this picture. Even more so, the explicit expressions for the isotropic case~\eqref{Re-F-iso},~\eqref{Im-F-iso} are not analytic because of a branch cut on the real $x^2$ axis that may be seen from~\eqref{F}, which although introduces a singularity, it does not give a straightforward pole structure wherein to identify ``particles''. Indeed, the branch cut singularity is a reflection that the intermediate dislon states can decay to (physical) on-shell phonons, and thus we do not expect those states to be able to survive  {for} arbitrarily long times.}

{It is therefore useful to distinguish the role of dislons as physically-propagating objects from their role as scatterers. While both are determined by the same exact propagator~\eqref{disl-prop-ex}, the values of $x$ in~\eqref{F} that become relevant to observables are starkly different:  {in the former, $x$ is fixed by the dispersion relation, whereas in the latter $x$ is a function of the angle at which the scattering phonon is incident on the dislocation.} }

\subsubsection{{Dislons as particles: evanescent and propagating modes}}

{To find the modes of propagation for dislons as point-like particles, we can use the ``usual'' on-shell condition to find the physically-propagating modes $S^{-1} = 0$, even with the explicit non-analytic expressions~\eqref{Re-F-iso},~\eqref{Im-F-iso}, provided they are analytically-continued for arbitrary complex numbers $x$ by re-deriving the result from~\eqref{F}. For simplicity, we proceed in the isotropic case, although we note that calculating the expressions from~\eqref{F} numerically is straightforward in the general, anisotropic, case.}

{In the isotropic case, this continuation is given by
\beq \label{F-iso-cont}
\begin{split}
F(x) &= \, \frac{g}{ 4\pi } \! \left[ \frac{1}{2} + \frac{3}{2 \gamma^4} + \left(1 - \frac{1}{\gamma^2} \right) c_T^2 x^2 \right. \\ & \quad \quad \quad \quad \left.   - (1 - c_T^4 x^4) \ln \! \left( \frac{ c_T^2 x^2 - 1 - i\epsilon}{c_T^2 x^2 - i\epsilon} \right)  \right. \\ & \quad \quad \quad \quad \left.  - \frac{(1 - \gamma^2 c_T^2 x^2)^2}{\gamma^4} \ln \! \left( \frac{ \gamma^2 c_T^2 x^2 - 1 - i\epsilon}{ \gamma^2 c_T^2 x^2 - i\epsilon} \right)   \right],
\end{split}
\eeq
where, for notational simplicity, we have introduced a dimensionless coupling constant $g$ to replace the role of the logarithm $\ln (\delta/\delta_0)$:
\beq
g \equiv \frac{4\pi}{2 (1+\gamma^{-4}) \ln(\delta/\delta_0) }.
\eeq}

{We must now find solutions to
\beq \label{disp-eq1}
\frac{2 (1 - \gamma^{-2})}{1+\gamma^{-4}} c_T^2 x^2 - \big(1 + F(x)\big) = 0,
\eeq
which can be searched for in terms of the dimensionless variable $z \equiv c_T x$. Note that in this setup, $z = c_T \kappa/\omega$, and  {consequently} we  {will} have a dispersion relation given by
\beq
\kappa(\omega) = z(g,\gamma) \frac{\omega}{c_T},
\eeq
meaning that the imaginary part of $z$ will determine the ``attenuation length'' of dislons at a fixed frequency as they propagate along the string. Conversely, one can define a dislon ``lifetime'' by writing $\omega(\kappa)$ and looking at the imaginary part of $z^{-1}$. }

{However, upon a closer look one finds a surprise as one ``turns on'' $g$ from zero: if one examines the function $F$, one sees that its imaginary part is nonzero around the ``free'' dispersion relation $\omega = \pm \kappa \sqrt{\Gamma/m}$, meaning the solution is no longer on the real $z$ axis. Moreover, the real $z^2$ axis with $0<z^2<1$ is precisely the branch cut of the logarithms in~\eqref{F-iso-cont}, meaning its imaginary part is discontinuous ``above'' and ``below'' this line (this is unambiguous if one refers to~\eqref{F}). Numerical analysis then shows that the propagating solution on the real $z^2$ axis in fact disappears for small (but nonzero) $g$, leaving  {the mentioned} branch cut singularity in its place.}

{It turns out that the solutions to~\eqref{disp-eq1} are found for negative $z^2$ as we increase $g$ from 0: this can happen if $F$ becomes more and more negative as $z^2$ approaches zero from the negative real $z^2$ axis. This is indeed the case: for negative $z^2$, $F$ is a monotonously decreasing function of $z^2$, starting from $F \to 0$ as $z^2 \to -\infty$ until it diverges when $z^2 \to 0$. Therefore, for arbitrary positive $g$ there exists a solution to $S^{-1} = 0$ with negative $z^2$, which means that the proportionality constant between $\omega$ and $\kappa$ is a purely imaginary number. These are evanescent waves, meaning that the excitations of the string (dislons) will always decay into phonons in a characteristic lifetime. The lifetime of these modes is given by
\beq
t_{\rm dislon}(\kappa) = |z_{\rm eva}|(g) (c_T \kappa)^{-1},
\eeq
where $|z_{\rm eva}|$, the value of $|z|$ associated to these evanescent waves, can be calculated numerically; it can take any value between 0 and $\infty$ as a monotonously increasing function of $g$.}

{Nonetheless, if we increase the value of $g$ sufficiently, then propagating solutions at real $z^2 > 1$ do re-appear. These are no longer affected by the branch cut, as the logarithms are real functions in this domain, because these states can no longer decay directly into physical phonons due to energy-momentum conservation. The critical value of $g$ so that these solutions appear is defined by the equation
\beq
g > g_c \equiv  \frac{4\pi \left(\dfrac{2 (1 - \gamma^{-2})}{1+ \gamma^{-4} } - 1 \right) }{\dfrac{3}{2} - \dfrac{1}{\gamma^2} + \dfrac{3}{2\gamma^4} - \dfrac{(1-\gamma^2)^2}{\gamma^4} \ln \left( \dfrac{\gamma^2-1}{\gamma^2} \right)   }.
\eeq
For instance, if $\gamma =2$, one finds $g_c \approx 3.44$.}

{Above this value of $g$, we will find propagating solutions for dislons that  {travel} with a ``renormalized'' speed 
\beq
c_{\rm dislon} = |z_{\rm prop}|^{-1}(g) c_T,
\eeq
where $|z_{\rm prop}|$ is the positive solution for $z$ to~\eqref{disp-eq1}. As in the evanescent case, $|z_{\rm prop}|$ is a monotonously increasing function of $g$, and can become arbitrarily large as $g \to \infty$, meaning that the speed at which these modes propagate can become arbitrarily small, and thus even relatively short-wavelength dislon excitations become low-energy particles in the theory. Therefore, in the large $g$ limit there can be a great number of excited dislon modes that cannot decay into physical phonons, and only serve as intermediate states in the quantum-mechanical path integral that represents the scattering amplitude of phonons. Indeed, unless additional couplings are introduced in the theory, these modes will effectively be decoupled from the rest of the theory.}

{We must note, however, that from a microscopical perspective $g$ cannot be arbitrarily large, as it is related to the logarithm of a division of cut-offs. At most, we could expect $g \sim 10$ if the cutoffs are related by $\delta \sim 2 \delta_0$.}

{Having discussed these aspects of dislons as particles, we now turn to the (perhaps) phenomenologically more interesting  {side} of the dispersion relation: interpreting the result when  {it is} placed in a phonon-to-phonon scattering amplitude.}

\subsubsection{{Dislons as scatterers}}

{While we have seen how propagating dislon modes disappear and re-emerge because of the dislon-phonon interaction, one can also interpret the result for the exact dislon propagator~\eqref{disl-prop-ex} from the point of view of the scattering amplitude~\eqref{scatt}. Indeed, we can write
\beq
\T = m E^*(\k;\iota) E(\k',\iota') \, \frac{i}{\hbar} S(k_3,\omega),
\eeq
where $S$ is fully determined by the dislon dispersion relation; in fact, it is basically the inverse of the operator that determines the exact equation of motion (in Fourier space) for dislons propagating along the dislocation line after integrating out the other degrees of freedom in the system (phonons).}

{This is a natural point to make contact with previous studies on phonon scattering by dislocations. It has long been recognized \cite{Ninomiya1974,Bitzek2004,Blaschke2019} that a classically described (i.e., non-quantum) moving dislocation will experience a drag force because of its interaction with phonons (a ``phonon wind''). Within the description that we have adopted in the present paper, this amounts to supplementing the string dynamics that follow from the action (\ref{action2}-\ref{action3}) with a phenomenological term to obtain}
\beq \label{B-pheno}
m \ddot X(s,t) + B \dot X(s,t) - \Gamma X''(s,t) = F(s,t),
\eeq
where $B$ is a phenomenological drag parameter and $F$ is the Peach-Koehler force. {In particular, in the context of}~\cite{Churochkin2016} {a first-order computation of phonon scattering in perturbation theory was carried out to determine the observable effects of this damping parameter $B$, leaving it as an adjustable quantity. Since it is a first-order computation, the dislon propagator connecting the ``external'' phonons (in a diagrammatic sense) is  {correspondingly} given by
\beq
S(\kappa,\omega) = \frac{-i \hbar}{- m \omega^2 - iB \omega + \Gamma \kappa^2}.
\eeq} 

{It is then natural to try and obtain a quantitative estimate for $B$ from our explicit results for $\Pi$, the dislon self-energy. If we interpret the real part of $F$ as a term that renormalizes the speed at which dislons propagate through the dislocation line at different wavelengths, then we may simply read off
\beq
- i B \omega = - i {\rm Im} \{\Pi(\kappa,\omega) \}.
\eeq
Because of the optical theorem, the imaginary part of $\Pi$ will be nonzero only for values of $ x = \kappa/\omega$ that allow for a decay to phonon states conserving momentum along the $\hat e_3$ direction, as well as frequency/energy $\omega$. This means that we can focus on the region where $|x| < c_T^{-1}$. To get an order-of-magnitude estimate, we may take $x=0$, where we have, in the isotropic case,
\beq
{\rm Im}\{ \Pi \} = \omega^2 \frac{\mu^2 b^2}{8 \rho c_T^4} \left(1 + \gamma^{-4} \right),
\eeq
implying
\beq \label{B-bound}
B \lesssim  \frac{1 + \gamma^{-4}}{8}  \rho b^2 \times \omega,
\eeq
where $\lesssim$ means that the phenomenological value of $B$ should be less than the RHS of~\eqref{B-bound}, depending on the angle of incidence of the phonon with respect to the dislocation line, but of the same order of magnitude.}

{While, on the one hand, this means that $B$ depends on the frequency of the incident phonon scattering off the dislocation line, this also provides a quantitative estimate that can be tested by comparing with experiments that intend to probe and characterize the equation of motion for dislocations phenomenologically, as with~\eqref{B-pheno}. }

{Finally, it is important to note that this estimate relied on using the exact dissipation rate computed from dislons propagating on an infinite string.  {However}, experiments testing this result may be sensitive to the length of the dislocation line $L$, which undoubtedly yields a different notion of dislon self-energy because momentum along the $\hat e_3$ axis is no longer conserved~\cite{Lund2019} as the exact equation of motion becomes infinitely coupled between the different modes of the string, and the identification of $B$ with the imaginary part of the inverse propagator needs to be revisited. In this case, one possibility for a direct identification would be to simply identify $B$ with the decay rate of a given dislon mode, probably corresponding to the first normal mode of the string.} {Additional effects to consider in order to make contact with experimental results should include a non-vanishing mean dislocation velocity, the effect of cubic and higher order terms in the phonon-dislon interaction and the effect of a finite temperature. Although it should be possible to tackle these phenomena within the formalism we present, doing so is outside the scope of the present work.} 

\subsubsection{{Resonances}}

{The preceding discussion,  {that is, the identification of a decay rate from the imaginary part of the self-energy $\Pi$, is tantamount to quantifying} the width of the resonance peak when a phonon scatters off a dislocation,  {as long as the coupling $g$ is ``weak'' ($g \ll 1$), and so} the ``free'' kinetic terms dominate.  {On the other hand,} at large coupling $g \gg 1$ the self-energy $\Pi$ becomes large and there is no obvious notion of a ``resonance'', since the virtual dislons in the amplitude will never be close to being ``on-shell'' in the sense of the free theory.  {In this situation, it wouldn't be possible to infer $B$ from a ``resonance'' peak in the phonon-to-phonon amplitude.} }

{Nonetheless, a further discussion of resonances makes  {perfect} sense in the small $g^2$ limit. In this limit, the dislon self-energy is negligible except for the (small) cut-off it provides to the on-shell divergence {of the free dislon propagator}. The peak will be located at an incidence angle to the dislocation of $\cos^2 \theta_p = (c_\iota k_3/\omega)^2 = c_\iota^2 m/\Gamma $. For transverse incident polarization, this corresponds to
\beq
\cos^2 \theta_{p,T} = \frac{1+\gamma^{-4}}{2(1- \gamma^{-2})}  > \frac12,
\eeq
which means that the cross-section will peak at a direction that is closer to $\hat e_3$ than to either $\hat e_1$ or $\hat e_2$. Presumably, this would give a larger thermal conductivity perpendicular to the dislocation line.} 

{This is disfavored by experiments~\cite{Sun2018}, and moreover, it is physically suspect from the microscopic point of view, where we would need a large value of $\ln (\delta/\delta_0)$.  This is not a sensible limit because it requires $\delta$ to be  {many} orders of magnitude greater than $\delta_0$, and from the microscopic point of view we expect (at most) 2-3 orders of magnitude (corresponding to $\ln (\delta/\delta_0) \sim 8$ as an upper bound).}

{For completeness, we note that for  {longitudinal incident} polarization, the peak would be at angles corresponding to
\beq
\cos^2 \theta_{p,L} = \gamma^2 \frac{1+\gamma^{-4}}{2(1- \gamma^{-2})}  > 1,
\eeq
which is geometrically impossible to attain. Only at the lowest possible value of $\gamma$, with $\lambda \ll \mu$, one might be able to observe a resonance in the limit where the incident phonon is collinear to the dislocation line.}

\begin{figure}
\includegraphics[width=0.5\textwidth]{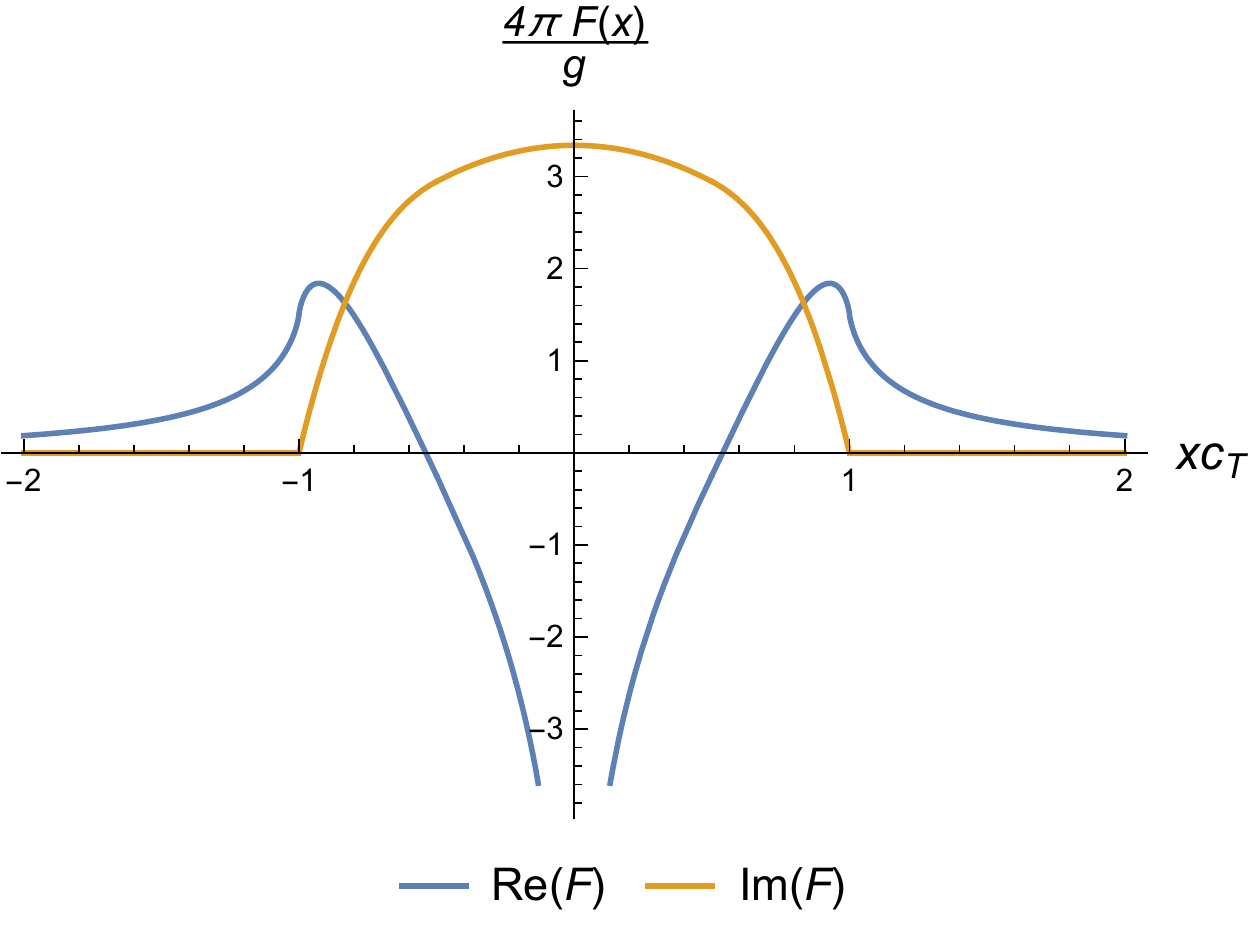}
\caption{Real and Imaginary parts of the function $F$, which determines the self-energy of the dislons through $\Pi = m \omega^2 F$ in the isotropic case. They do not vanish simultaneously. The plots were generated setting $\gamma=2$.}
\label{fig:F-form}
\end{figure}

{The other limit, $g \gg 1$, shows no resonances in the scattering cross-section, as the imaginary and real parts of $F$ never approach zero simultaneously (at least in the isotropic case; see Figure~\ref{fig:F-form}). In contrast, this limit leads to a suppression of the scattering amplitude when the incident phonon becomes perpendicular to the dislocation line, because the function $F(x)$ diverges logarithmically as $x \to 0$.} 

{This last particularity is due to the fact that the dislocation segment under consideration is infinite, but pinned in its endpoints. The infinity of its extension implies an extra symmetry, which then gives momentum conservation. Then, scattering of phonons with $k_3 = 0$ by the dislocation line should excite the $\kappa=0$ mode of the string, which would correspond to a uniform translation. But this excitation is not possible because the dislocation is pinned: the  {boundary conditions} of the theory forbid such process. In contrast to this, a finite dislocation segment does not imply a vanishing phonon cross-section at $x=0$ because $k_3$ need not be conserved, and as such, other modes (with nonzero wavenumber) can be excited on the string.}

\subsection{Thermal equilibrium} \label{sec:th-equil}

{As a check on the consistency of the formalism, we verify} that the previous coupling automatically provides a mechanism with which phonons and dislons, the excitations on the dislocation line, reach thermal equilibrium. 

Let us consider a single dislocation inside of a solid set in an environment at temperature $T$. Far away from the dislocation, the expectation value for the number of phonons is given by the Bose-Einstein distribution
\beq 
\langle n_{\k,\iota} \rangle = \frac{1}{e^{\hbar \omega_{\iota}(\k) /k_B T} - 1 },
\eeq
which in turn sources the states that will later scatter with the dislocation line. In thermal equilibrium, therefore, we expect that the dislon distribution $f_{\rm d}$ at wavenumber $\kappa$ be such that {the rate at which a dislocation mode decays,  {denoted by $P_{\rm d-decay}(\kappa)$,} is related to the probability per unit of time of a phonon being ``absorbed'' into a dislocation excitation $P_{{\rm ph} \to {\rm d}}$ by:}
\beq \label{equilibrium}
P_{\rm d-decay}(\kappa) f_{\rm d}(\kappa,T) = \sum_\iota \int_\k \frac{P_{{\rm ph} \to {\rm d} }(\k,\iota;\kappa) }{e^{\hbar \omega_{\iota}(\k) /k_B T} - 1 }.
\eeq
However, dislons decay precisely into phonons {through rates $P_{ {\rm d} \to {\rm ph} }$} that satisfy
\beq
P_{\rm d-decay}(\kappa) = \sum_\iota \int_\k P_{ {\rm d} \to {\rm ph} }(\kappa; \k, \iota),
\eeq
and moreover, both processes (${\rm ph} \to {\rm d}$ and viceversa) must contain $\delta (\omega_\iota(\k) - \omega_\kappa )$ as an overall factor. This implies that~\eqref{equilibrium} is actually
\beq
\begin{split}
& f_{\rm d}(\kappa,T) \sum_\iota \int d\Omega  P_{ {\rm d} \to {\rm ph} }(\kappa; \omega_\kappa/c_\iota(\hat{k}), \iota)  \\ &= \frac{1}{e^{\hbar \omega_\kappa /k_B T} - 1 } \sum_\iota \int d\Omega P_{{\rm ph} \to {\rm d} }(\hat{k} \omega_\kappa/c_\iota(\hat{k}) ,\iota;\kappa)
\end{split},
\eeq
and therefore, because the amplitudes that give rise to both  {probabilities/rates of decay under the angular integral sign ($P_{ {\rm d} \to {\rm ph} }$ and $P_{{\rm ph} \to {\rm d} }$)} have the same absolute value (they are mapped onto each other by time reversal,  {which is a symmetry of this model}), we find
\beq
 f_{\rm d}(\kappa,T) = \frac{1}{e^{\hbar \omega_\kappa /k_B T} - 1 }.
\eeq

This implies that if the solid is in an environment with temperature $T$, then the modes of the strings (dislocations) will also feel the same temperature.

\section{Implications on thermal transport}
\label{sec:thtr}

{We now turn to examining how the scattering mechanism provided by dislocations affects energy transport in a solid. In particular, we focus on thermal transport through phonons, and set the groundwork for the computation of thermal conductivities in a solid threaded by highly-oriented dislocations. In this section, we start by discussing conventional cross-sections in a scattering picture, and then move on to consider the lifetime of phonons participating in thermal transport. Finally, we compare with the original work of Klemens~\cite{Klemens1955} and Carruthers~\cite{Carruthers1959} and study the anisotropy in thermal transport that arises due to a large number of long dislocations threading the solid.}

\subsection{The scattering of a phonon by a dislocation: cross sections} \label{sec:csecs}

{From~\eqref{scatt}, the differential scattering cross section from mode $(\k, \iota)$ to mode $(\k', \iota')$ is given directly by taking the absolute value squared of the scattering amplitude $\T$, integrating over the length of the wave-vector $k = |\k|$, and divide by the incident flux $ v_\iota^g(\hat k)/V $ times the norm of the incident state $V$ (where $V$ is the volume of the elastic continuum, and $v_{\iota}^g( \hat k) = |d\omega_\iota / d\k|$ is the group velocity of sound waves in the elastic medium). One obtains}
\beq \label{diff-cross}
\begin{split}
\frac{d \sigma_{\iota \iota'} }{d\Omega} = \, & \frac{L  \delta (k_3 - k_3')}{2\pi  c_{\iota'}^5(\hat{k}') c_\iota^2(\hat{k}) v_{\iota}^g( \hat k) } \left( \frac{N^2 b^2}{2 \rho m} \right)^2 \\ & \times \frac{ |\hat k_l {\bf M}_{kl} \varepsilon_\iota(\k)_k |^2 |\hat{k}'_{l'} {\bf M}_{k'l'} \varepsilon_{\iota'}(\k')_{k'} |^2 }{ \left| \frac{\Gamma k_3^2}{m \omega^2} -\left( 1 + F \! \left( \frac{ k_3}{\omega} \right)  \right)  \right|^2 },
\end{split}
\eeq
{where $L$ is the length of the dislocation, which we take to be large so that the approximation $L \sim (2\pi) \delta(\kappa - \kappa) = (2\pi) \delta(0) $ is justified.}

 {Integrating over the possible outgoing states, i.e., over the relative angle $d\Omega$ between $\k$ and $\k'$}, the total cross section for the mode $(\k, \iota)$ may be written as
\beq
\begin{split}
\sigma_\iota(\k) = \, & \frac{ L}{2  c_\iota^2(\hat{k}) v_{\iota}^g( \hat k) } \left( \frac{N^2 b^2}{2 \rho m} \right)^2 \frac{ |\hat k_l {\bf M}_{kl} \varepsilon_\iota(\k)_k |^2}{ \left| \frac{\Gamma k_3^2}{m \omega^2} -\left( 1 + F \! \left( \frac{ k_3}{\omega} \right)  \right)  \right|^2 } \\ & \times \sum_{\iota'} \int_{-1}^1    du (1 - u^2)  \frac{  |\varepsilon_{\iota'}^{xy}(u) |^2 }{c_{\iota'}^5(u)} \delta \! \left( k_3 -  \frac{\omega u}{c_{\iota'}(u)} \right), 
\end{split}
\eeq
and averaging over dislocation orientations (Burgers vector) in the  {$\hat e_1 - \hat e_2$} plane one obtains
\beq \label{sigma-iota}
\begin{split}
\bar{ \sigma }_\iota(\k) =& \, \frac{ L}{4 \omega_\iota(\k) c_\iota^2(\hat{k}) v_{\iota}^g( \hat k) } \left( \frac{N^2 b^2}{2 \rho m} \right)^2 \frac{ |\varepsilon_{\iota}^{xy}(\cos \theta) |^2 \sin^2\theta }{ \left| \frac{\Gamma k_3^2}{m \omega^2} -\left( 1 + F \! \left( \frac{ k_3}{\omega} \right)  \right)  \right|^2 } \\ 
& \times \sum_{\iota'}  \int_{-1}^1    du (1 - u^2)  \frac{  |\varepsilon_{\iota'}^{xy}(u) |^2 }{c_{\iota'}^5(u)} \left| \frac{c_{\iota'}^2(u) }{c_{\iota'}(u) - u c'_{\iota'}(u) } \right| \\ & \quad \quad \quad \quad \times \delta \! \left( u -  u_{\iota'}(k_3/\omega) \right), 
\end{split}
\eeq
where $\cos \theta = \hat k \cdot \hat e_3$, $u_\iota(x)$ is defined as the solution to the equation $u = x c_\iota(u)$, and $c'_\iota(u) \equiv d c_\iota(u)/du$.

{{We can proceed further without overcomplicating the expressions if we assume an isotropic elastic continuum, because here we only have two sound speeds, $c_T$ ($\times 2$) and $c_L$, corresponding to transverse and longitudinal polarizations, that do not depend on the direction of propagation $\theta$.} Moreover, the sum over transverse polarizations can be evaluated (for an isotropic medium) to 
\beq
\sum_{\iota = T_1,T_2} |\varepsilon_{\iota}^{xy}(u) |^2 = 1 + u^2,
\eeq
whereas for longitudinal polarization we have 
\beq
|\varepsilon_{L}^{xy}(u) |^2 = 1 - u^2.
\eeq
The integral over $u$ is now straightforward, as the Dirac delta becomes $\delta(u - c_{\iota'} k_3/\omega)$. It gives (in a strictly isotropic elastic continuum)
\beq
\begin{split}
& \sum_{\iota'}  \int_{-1}^1    du (1 - u^2)  \frac{  |\varepsilon_{\iota'}^{xy}(u) |^2 }{c_{\iota'}^5(u)} \left| \frac{c_{\iota'}^2(u) }{c_{\iota'}(u) - u c'_{\iota'}(u) } \right| \\ & \quad \quad \quad \quad \times \delta \! \left( u -  u_{\iota'}(k_3/\omega) \right) \\
& \quad \quad = \frac{1}{c_T^4} \left( 1 - \left( \frac{c_T k_3}{\omega} \right)^4 \right) \Theta \left( 1 - \left| \frac{c_T k_3}{\omega} \right| \right) \\ & \quad \quad \quad + \frac{1}{c_L^4} \left( 1 -  \left( \frac{c_L k_3}{\omega} \right)^2 \right)^2 \Theta \left( 1 - \left| \frac{c_L k_3}{\omega} \right| \right) \\
& \quad \quad \equiv \frac{1}{c_T^4} I(k_3/\omega).
\end{split}
\eeq
For physical phonons, the first Heaviside function is always one because $\omega = c_\iota k \implies |c_T k_3/\omega| < 1$. To cast everything in terms of dimensionless functions, we can define
\beq
A(k_3/\omega) \equiv \left| \frac{\Gamma k_3^2}{m \omega^2} - \left( 1 + F \! \left( \frac{ k_3}{\omega} \right)  \right)  \right|^2,
\eeq
which captures the polarization-independent contribution  {that depends} on $k_3/\omega$. We can also work through the $N^2 b^2/(2\rho m)$ factor in the isotropic limit, where $N = \mu$ and $m = \rho b^2 (1 + \gamma^{-4} ) \ln(\delta/\delta_0)/(4\pi)$, giving
\beq
\left( \frac{N^2 b^2}{2 \rho m} \right)^2 = c_T^8 \left( \frac{2\pi}{ (1 + \gamma^{-4}) \ln (\delta/\delta_0) } \right)^2 \equiv c_T^8 g^2,
\eeq
where, as before, we have introduced the dimensionless coupling constant $g$ for notational simplicity. {In this form, the cross-sections of a phonon scattering by a single dislocation read
\beq
\sigma_T(\k) = \frac{L c_T g^2}{8 \omega_T(\k)} (1 - \cos^4 (\theta) )  \frac{I(k_3/\omega_T(\k))}{A(k_3/\omega_T(\k))}
\eeq
and
\beq
\sigma_L(\k) = \frac{L c_T g^2}{4 \gamma^3 \omega_L(\k)} \sin^4(\theta) \frac{I(k_3/\omega_L(\k))}{A(k_3/\omega_L(\k))},
\eeq
where we have averaged over the two polarizations in the transverse case.
}}

\subsection{The scattering of a phonon by a dislocation: lifetimes} \label{sec:ltimes}

{We now turn to the task of estimating the phonon lifetime in thermal transport due to scattering by dislocations. For simplicity, we shall assume that the elastic continuum is isotropic, {and will consider $\Lambda_d$ parallel dislocations per unit area.} This is a slightly different calculation to that of the cross-section, because in writing down an equation for the evolution of the expected occupancy of mode $(\k;\iota)$, we need to include transition probabilities both \textit{from} and \textit{to} any other mode in the theory.} {Our goal will be to calculate the single-mode phonon decay rates $\tau^{-1}_\iota(\k)$, so that they may later be used to compute the thermal conductivity tensor using the relation~\cite{Soto2016}
\beq
\begin{split}
K_{ij} = \sum_{\iota} \! \int \!\! \frac{d^3 k}{(2\pi)^3} & \frac{ e^{\hbar \omega_\iota(\k)/k_B T}}{\left(e^{\hbar \omega_\iota(\k)/k_B T } - 1\right)^2} \\ & \times \frac{\hbar \omega_\iota(\k) \tau_\iota(\k)}{ k_B T^2 }  [{\bf v}_\iota(\k)]_i [{\bf v}_\iota(\k)]_j,
\end{split} \label{th-cond}
\eeq
where ${\bf v}_\iota(\k)$ is the phonon velocity of propagation for the mode $(\k;\iota)$. In a sense, this is a relaxation time approximation, because equation~\eqref{th-cond} assumes that all transport phenomena can be described through a single phonon lifetime $\tau_\iota(\k)$ for each mode separately. } 

\begin{figure*} 
\includegraphics[width=0.8\textwidth]{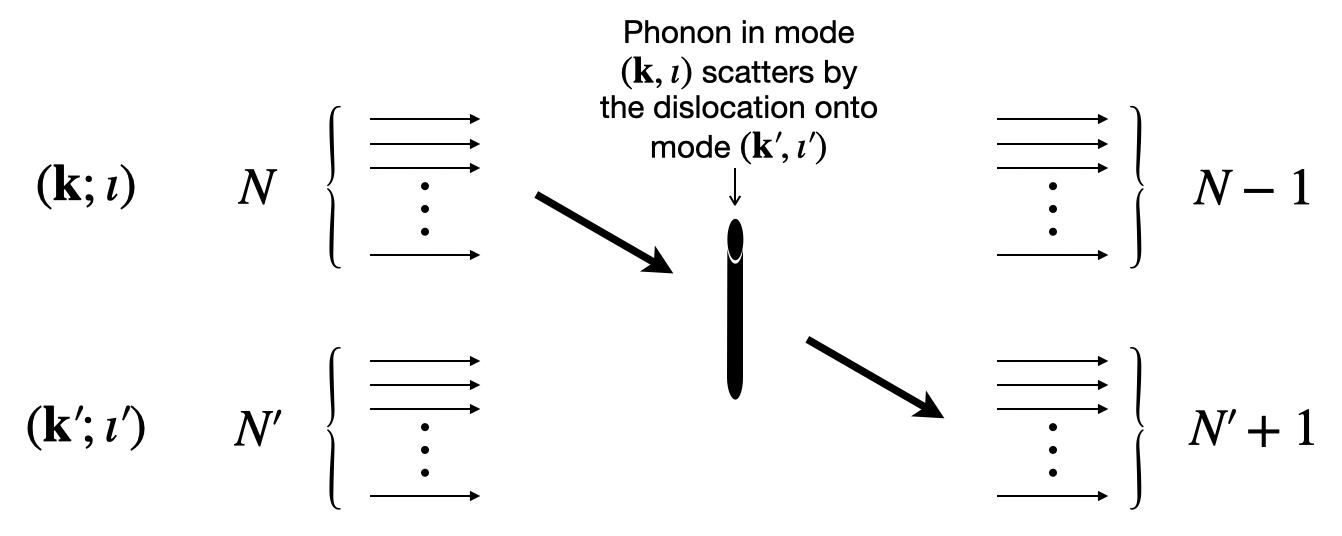}
\caption{Diagrammatic illustration of the process that gives the rate at which phonons populating the mode $(\k;\iota)$ transition to $(\k';\iota')$.} \label{fig:lifetime}
\end{figure*}

{The scattering processes that contribute to this phonon lifetime may be illustrated as in Figure~\ref{fig:lifetime}: one of the phonons of mode $(\k;\iota)$ scatters off the dislocation line, and goes into the mode $(\k',\iota')$. Out of these individual processes, we want to first determine the rates of transition between the different modes, and then write down the full lifetime of mode $(\k;\iota)$ by subtracting the rate at which phonons are created in this mode with the rate at which they decay.}

{ The derivation of the lifetimes proceeds as follows: since the interaction under consideration couples an ingoing $(\k,\iota)$ mode with an outgoing $(\k',\iota')$, the relevant amplitude admits the following schematic representation in terms of harmonic oscillator ladder operators ($a$ and $a'$ for the modes $(\k;\iota)$ and $(\k';\iota')$, respectively)
\beq
\braket{f | \T_{1-1} a (a')^{\dagger} | i}
\eeq 
with 
\beq
\ket{i} = \frac{(a^\dagger)^N}{N!} \frac{((a')^\dagger)^{N'}}{N'!} \ket{0} = \ket{N,N'}
\eeq 
and 
\beq
\ket{f} = \frac{(a^\dagger)^{N-1}}{(N-1)!} \frac{((a')^\dagger)^{N'+1}}{(N'+1)!} \ket{0} = \ket{N-1,N'+1},
\eeq 
 {where $\T_{1-1}$} represents the one-to-one particle transition amplitude:  {it is essentially a placeholder for the phonon-to-phonon scattering amplitude in~\eqref{scatt}}. Standard algebra in quantum mechanics then gives
\beq
\braket{f | \T_{1-1} a (a')^{\dagger} | i} = \T_{1-1} \sqrt{N(N'+1)},
\eeq
implying that the transition rate is proportional to $|\T_{1-1}|^2 N (N'+1)$. Conversely, the rate of transition from mode $(\k',\iota')$ to $(\k;\iota)$ is proportional to $|\T_{1-1}|^2 N' (N + 1)$.} 

{Let us stress that, in the above discussion, we have assumed that the transitions of interest involve only one scattering process at the same time, but, as opposed to what one would do in leading-order perturbation theory, we keep the full one-to-one phonon interaction amplitude, which accounts for all the scattering dynamics of a single phonon. Including simultaneous transition processes is feasible within the framework presented in this work, but it falls outside the scope of the present approximation, in which only single-mode lifetimes are considered.}

{If we now assume the number distribution $N$ can be written as an equilibrium distribution $N_0$ plus a deviation $n$, i.e., $N = N_0 + n$, then the time derivative of the occupancy of mode $(\k;\iota)$ is proportional to
\beq
\begin{split}
\sum_{(\k';\iota')} |\T|^2 \left( (N+1)N' - N(N'+1) \right) &= |\T|^2 (N' - N) \\
&= |\T|^2 (n' - n),
\end{split}
\eeq
 {where we have assumed that the equilibrium distribution $N_0$ is the same for all phonon modes. This is indeed the case if said equilibrium distribution is the Bose-Einstein distribution.}
In particular, if we have $\Lambda_d$ parallel dislocations per unit area, this means that the relaxation time for mode $(\k,\iota)$, which can be written as $\tau_\iota(\k)^{-1} = - \dot n_{\iota,\k}/n_{\iota,\k} $, is given by
\beq \label{tau-def-cross}
\tau_\iota(\k)^{-1} = \frac{\Lambda_d v_\iota^g(\hat k)}{L} \sum_{\iota'} \int d\Omega \frac{d \sigma_{\iota \iota'}}{d\Omega} \left( 1 - \frac{n_{\iota'}}{n_\iota} \right),
\eeq
where $d\sigma_{\iota \iota'}/d\Omega$ is given by~\eqref{diff-cross},  {the differential cross section in vacuum}. }

{In the presence of a temperature gradient, the out-of-equilibrium occupation numbers $n_\iota$, $n_{\iota'}$ should reflect the fact that heat is being transported along a fixed direction. Following the works of Klemens~\cite{Klemens1955} and Carruthers~\cite{Carruthers1959}, we use the estimate  {
\beq \label{n-assumption}
n_{\iota,\k} \propto {\hat{k} \cdot \nabla T},
\eeq 
in the spirit that the out-of-equilibrium distribution will imply a heat current in the direction defined by $\nabla T$.} We leave the examination of  {this assumption} from a more modern perspective of thermal transport using linear response coefficients~\cite{Luttinger1964} in thermal quantum field theory (as recently suggested~\cite{Li2018}) for future work.}

{There are two main cases of interest: thermal transport parallel to the dislocation lines and perpendicular to them. The first case is a direct extension of the cross-sections we computed in the previous section,  {as 
\beq 
\nabla T \parallel \hat e_3 \implies \frac{n_{\iota'}}{n_\iota} = d_{\iota' \iota} \frac{\cos (\theta_{\rm out}) }{\cos (\theta_{\rm in})}, 
\eeq
where $d_{\iota \iota'}$ is unity if $\iota = \iota'$ (because the proportionality constant in~\eqref{n-assumption} is the same), and a number to be determined if $\iota \neq \iota'$, satisfying $d_{\iota \iota'} = (d_{\iota' \iota})^{-1}$.} Because both $k_3$ and $\omega$ are conserved for transverse-to-transverse as well as for longitudinal-to-longitudinal scattering, in these situations this factor is equal to one, and therefore these processes do not contribute to the phonon lifetime. 

 {In principle,} transverse to longitudinal and vice-versa processes  {could contribute. Note that} because both ingoing and outgoing scattering angles satisfy $\cos \theta = k_3/|\k| = c_\iota k_3/\omega$, the conservation of $k_3/\omega$ implies that $ c_\iota^{-1}  \cos (\theta)$ is also conserved amongst ingoing and outgoing modes. Therefore,
\beq \label{T-to-L}
\frac{\cos \theta_T}{c_T} = \frac{ \cos \theta_L}{c_L} \implies \gamma \cos \theta_T = \cos \theta_L,
\eeq
meaning that longitudinal polarization can always scatter to transverse polarization, but for some angles transversely polarized phonons cannot scatter onto longitudinal modes.}

 {Now comes a crucial observation: since we expect a steady current to be held in place, condition which is part of the definition of $n_{\iota,\k}$, all phonon lifetimes must be positive (if they were negative, it means that one particular mode continues to receive phonons from another mode perpetually). Since the kinematics of the phonon-to-phonon scattering process fix the angle of the outgoing phonon relative to the dislocation line $\theta$, the quotient $\cos (\theta_{\rm out})/\cos (\theta_{\rm in})$ can be either $\gamma$ or $1/\gamma$, and therefore the sign of $\tau_{\iota}(\k)^{-1}$ is fixed by $1 - n_{\iota'}/n_\iota$. However, because $\tau^{-1}_\iota$ is positive we must have
\beq
\frac{n_T}{n_L} = \frac{d_{TL}}{\gamma} \leq 1,
\eeq
for the longitudinal-to-transverse transition ratio, and
\beq
\frac{n_L}{n_T} =  d_{LT} \gamma \leq 1
\eeq
for the transverse-to-longitudinal ratio. This implies that $d_{TL} = \gamma = d_{LT}^{-1}$, because otherwise the kinematically allowed processes would drive the thermal current out of its steady state.}

 {Therefore, we have that
\beq \label{tau-T-par}
\tau_T^{\parallel}(\k)^{-1} = 0,
\eeq
\beq \label{tau-L-par}
\tau_L^{\parallel}(\k)^{-1} = 0.
\eeq
Therefore, thermal transport in the direction parallel to the dislocation can only be impeded by scattering mechanisms that are not due to dislocations, at least directly.}

{The other case of interest is to take the temperature gradient perpendicular to the dislocation line. For definiteness, we take the temperature gradient to be oriented along a line on the $\hat e_1 - \hat e_2$ plane, defined by an angle $\phi_\nabla$
\beq
\nabla T \parallel \cos (\phi_\nabla) \hat e_1 + \sin (\phi_\nabla) \hat e_2,
\eeq 
so that
\beq
\hat k' \cdot \nabla T \propto \sin(\theta') \left( \cos(\phi') \cos (\phi_\nabla) + \sin(\phi') \sin(\phi_\nabla) \right).
\eeq
with $\phi'$ the azimuthal angle of the outgoing phonon. Now, note that the $\phi'$-dependent piece in~\eqref{diff-cross} is given entirely by $|\hat k'_l {\bf M}_{kl} \varepsilon_{\iota'}(\k')_k |^2$, which, after a brief inspection, can be shown to involve an even number of trigonometric functions $\sin (\phi')$, $\cos (\phi)'$ as factors in the integrand. Since the integral over a full period of an odd power of trigonometric functions vanishes, we conclude that the $n_{\iota'}/n_{\iota}$ term does not contribute to the phonon lifetime. Therefore, the phonon lifetime for a temperature gradient perpendicular to the dislocation line is given by $\tau^{\perp}_\iota(\k)^{-1} = v_\iota^g(\hat k) \Lambda_d \sigma_\iota(\k)/L$. Explicitly,
\beq \label{tau-T-per}
\tau_T^{\perp}(\k)^{-1} = \frac{\Lambda_d c_T^2 g^2}{8 \omega_T(\k)} (1 - \cos^4(\theta))  \frac{I(\cos (\theta)/c_T)}{A(\cos (\theta)/c_T)},
\eeq
and
\beq \label{tau-L-per}
\tau_L^{\perp}(\k)^{-1} = \frac{\Lambda_d c_T^2 g^2}{4 \gamma^2 \omega_L(\k)} \sin^4(\theta)  \frac{I(\cos (\theta)/c_L)}{A(\cos (\theta)/c_L)}\, .
\eeq
Eqs. (\ref{tau-T-par}-\ref{tau-L-par}) and (\ref{tau-T-per}-\ref{tau-L-per}) constitute our results for the phonon lifetimes in an isotropic solid threaded by infinitely long dislocations along the $\hat e_3$ axis.}

{This model has one free parameter, given by the short- and long-distance cutoff lengths through $\ln (\delta/\delta_0)$, that appear in the theory when we idealize the dislocation as a string. Equivalently, we can take $g$ to be the free parameter in this description. All other quantities can be determined from macroscopic measurements of the elastic continuum, which makes the theory rather appealing in the sense that it is not overly sensitive to the microscopic constituents of the dislocation line.}

\subsection{A comparison with Klemens' and Carruthers' models}
\label{sec:KC}

{
At this point, it becomes paramount to compare these result with previous models for phonon scattering by dislocations. The Carruthers model~\cite{Carruthers1959}, after several approximations including considering a simple cubic lattice, and only considering incident phonons perpendicular to the dislocation (which is the incident direction of maximum scattering in that model), gives a relaxation time of
\beq
\tau^{\rm Carruthers}(\k)^{-1} = \frac{1}{3} |\k| \Lambda_d b^2 G^2 c_s \left[ \ln \left( b \sqrt{\Lambda_d} \right) \right]^2,
\eeq
where $G$ is the Gr\"uneisen parameter, and $c_s$ is the average sound speed in the material.
}

{
Klemens' model~\cite{Klemens1955}, which historically was introduced earlier, gives
\beq
\tau_{\iota}^{\rm Klemens}(\k)^{-1}_{\rm strain \, field} \propto \omega_\iota(\k) \Lambda_d b^2  G^2
\eeq
where the proportionality constant is an $\mathcal{O}(1)$ number that depends on the ratio of edge and screw dislocation densities, as well as on the Poisson ratio. This model also provides a phonon-dislocation scattering contribution from the cores of dislocations, which may be approximated as
\beq
\tau_\iota^{\rm Klemens}(\k)^{-1}_{\rm core} = \Lambda_d V_{a}^{4/3}  \omega_\iota(\k)^3/c_s^2,
\eeq
where $V_a$ is the volume per atom in the solid.
We can now compare these results with our expressions for the  {phonon lifetimes, in the case where} $\Lambda_d/L$ is the number of dislocations per unit of volume  {in a highly oriented array} (assuming the long dislocations thread the elastic continuum from side to side). 
}

{Let us start by examining the strength of the scattering. Quick inspection of our results (\ref{tau-T-per}-\ref{tau-L-per}) shows that in our model the phonon lifetime scales as 
\beq
\tau^{-1} \propto \Lambda_d c_s^2 \omega^{-1 }
\eeq
at fixed $\theta$, with the other factors being of $\mathcal{O}(1)$. {The fact that the scattering cross section for phonon scattering by a dynamically responding,  infinitely long, dislocation scales like the inverse of the phonon frequency goes back to early results of Eshelby and Nabarro~\cite{Eshelby1949,Nabarro1951,Maurel2004}}. The similarities and differences with Klemens' and Carruthers' models are evident at this point:
\begin{enumerate}
\item All models (even though Klemens' and Caurruthers' results we have shown here do not make this explicit) have a vanishing phonon decay rate at $\theta = 0, \pi$, i.e., when the phonon is incident parallel to the dislocation line, favoring thermal transport in this direction over the others.
\item All models have a linear dependence on the dislocation density, with the observation that Carruthers' model has an additional logarithmic sensitivity to the dislocation density because of how the strain field is modeled. This makes the interaction strength of Carruthers' model generically stronger than Klemens'.
\item The other parameters that control the magnitude of the phonon lifetime are $c_s$, the sound speed in the material, and $b$, the dislocation's Burgers vector. Incidentally, our model is insensitive to the value of the Burgers vector, being only dependent on the macroscopic parameters $c_s$ and $\gamma$.
\item In stark contrast to what both Klemens' and Carruthers' models predict, the phonon lifetime in our model is larger at smaller frequencies, depending on the phonon energy as $\omega^{-1}$ over the range of frequencies where the infinite dislocation line approximation holds $kL \gg 1$. In particular, this means that the thermal transport anisotropy induced by dislocations will become stronger at lower temperatures relative to Carruthers' and Klemens' models.
\end{enumerate}
This last point may prove to be crucial in explaining the low-temperature dependence of the thermal conductivity in a material threaded by dislocations from side to side, as has been recently observed by Sun et al. in thin InN films~\cite{Sun2018}, an effect that is not captured by earlier models. This will be explored quantitatively in upcoming work.
}

\subsection{Thermal transport anisotropy}
\label{sec:anisotropy}

{Note that one clear advantage of our result is that the angular dependence of the phonon lifetime on the polar angle $\theta$ is explicit, and therefore we can compute estimates for the anisotropy in thermal conductivity quantitatively.  {We} proceed in the isotropic case, where we have explicit expressions for the scattering cross-sections  {and lifetimes}.  {At each fixed frequency $\omega$, the differential thermal conductivity tensor $d K_{ij}$, i.e., the contributions that the full thermal conductivity tensor $K_{ij}$ receives from modes with single-phonon energies of $\hbar \omega$, may be used to study the generation of thermal transport anisotropy at each energy scale. In particular, we can write}
\beq
d K_{ij} \propto d\omega   \sum_\iota \int \! d\Omega \, c_\iota^2 \hat k_i \hat k_j \tau_\iota(\k),
\eeq
where we have omitted other temperature- and energy-dependent factors. Furthermore, if we use that $\tau$ only depends on the direction of propagation through the angle $\theta$, one gets (now in matricial notation, where the first two rows/columns correspond to the $\hat e_1$, $\hat e_2$ directions and the third to $\hat e_3$)
\beq
\begin{split}
d {\bf K} &\propto d\omega   \sum_\iota c_\iota^2 \int_0^\pi d\theta \sin (\theta) \\ &   \times \begin{bmatrix} \frac{\sin^2 (\theta) }{2}  \tau^\perp_\iota(\omega, \theta) & & \\ & \frac{\sin^2 (\theta) }{2}  \tau^\perp_\iota(\omega, \theta) & \\ & & \cos^2(\theta) \tau^\parallel_\iota(\omega, \theta)
\end{bmatrix},
\end{split}
\eeq
in which we have made explicit that the decay rate $\tau^{-1}$ depends only on the phonon energy and on the angle between the direction of propagation with the dislocation line axis.
}

\begin{figure*}[t] 
\includegraphics[width=0.42\textwidth]{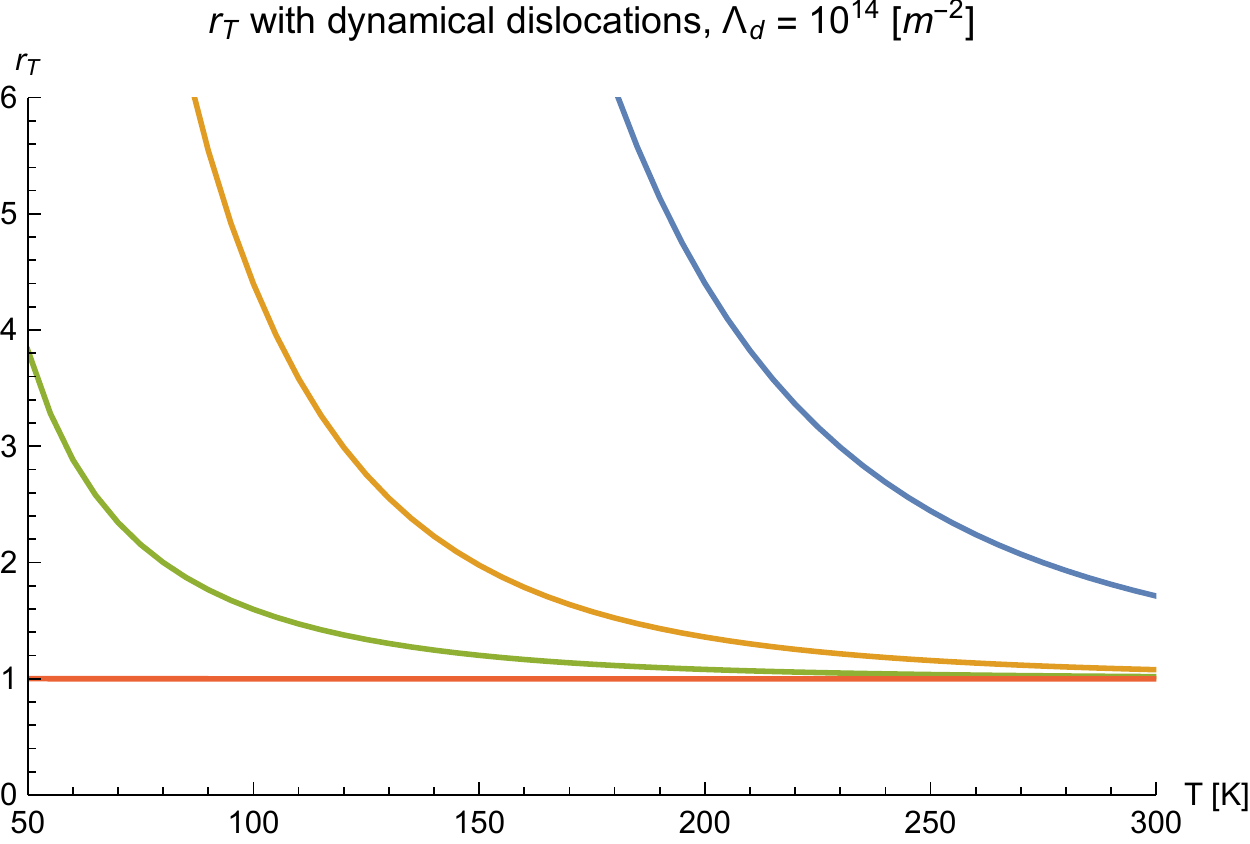}
\includegraphics[width=0.56\textwidth]{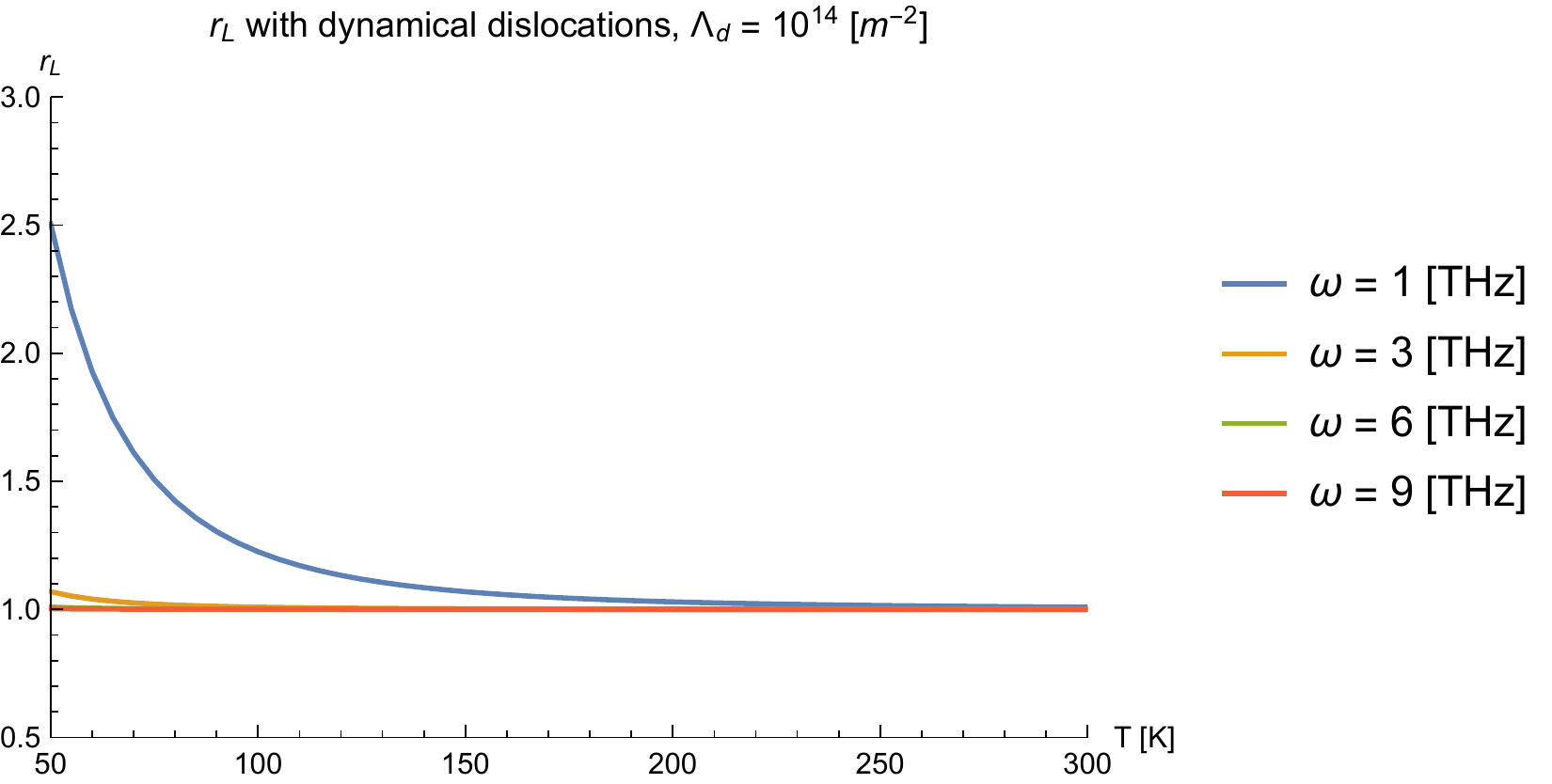}
\includegraphics[width=0.42\textwidth]{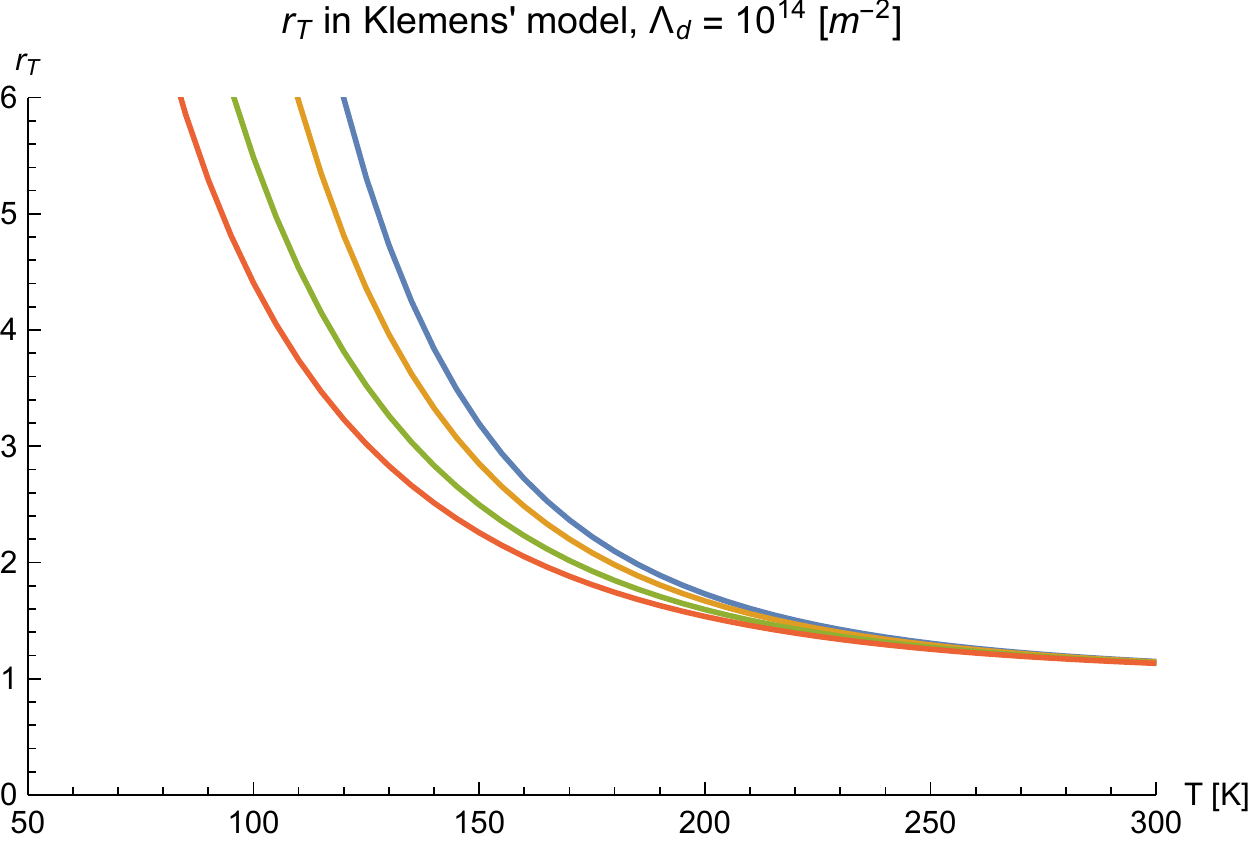}
\includegraphics[width=0.56\textwidth]{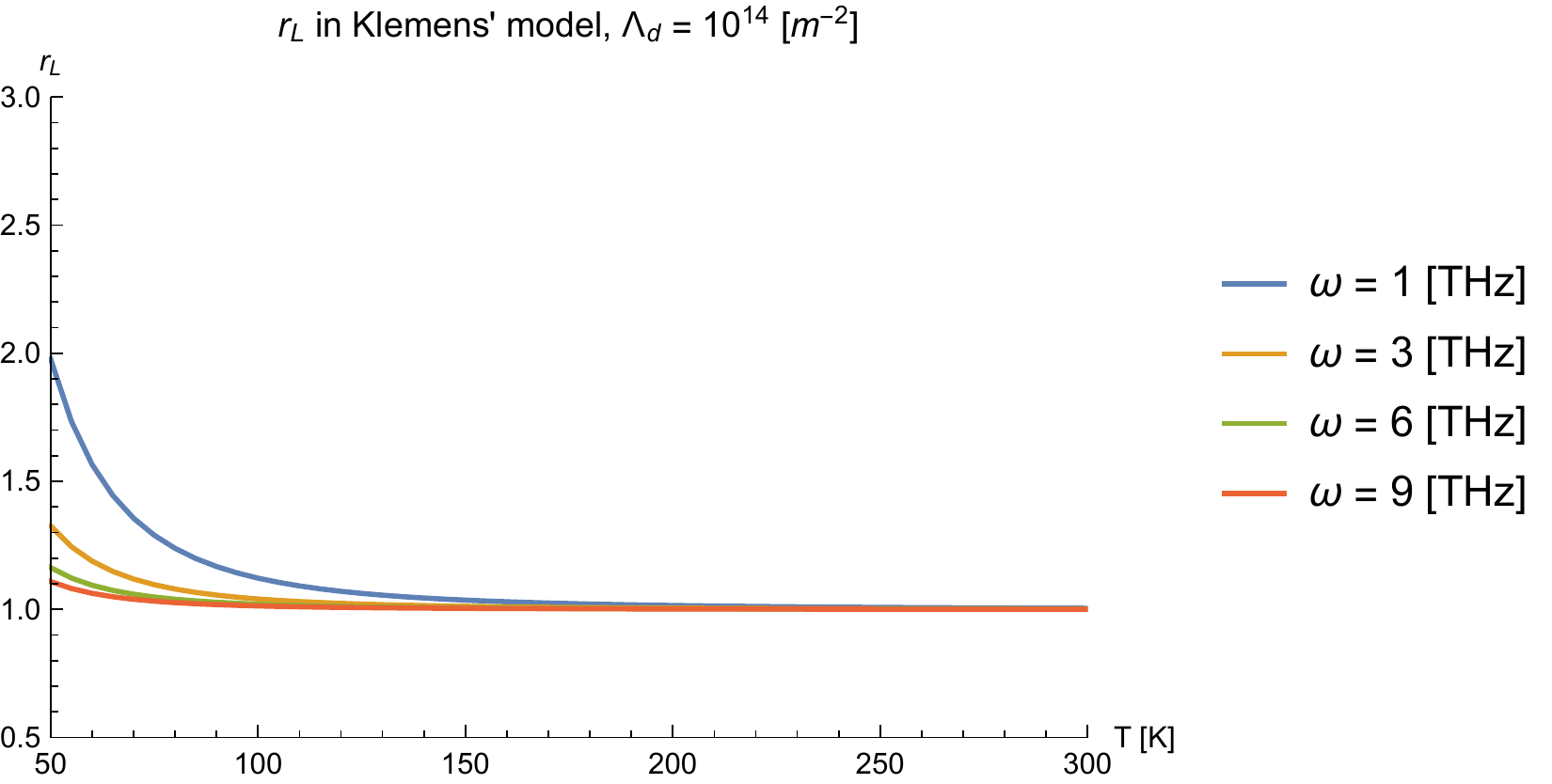}
\caption{{Differential anisotropy ratio $r_\iota$ as a function of temperature $T$ for different frequencies  $\omega$. {\it Upper panels:} Present work, based on a dynamical response of dislocations to phonons. {\it Lower panels:} Klemens model, based on a static response of dislocations to phonons. {\it Left-hand-side panels:} Transverse polarization. {\it Right-hand-side panels:} Longitudinal polarization.  The plots were calculated for  {$\gamma=2$, $g = 3$, $G = 2$, $b = 3.5 \times 10^{-10}$ m,} a dislocation density of $\Lambda_d = 10^{14}$ m${}^{-2}$, and the  {intrinsic phonon lifetimes parametrizations} of Ge~\cite{Asen1997} . There is a significantly different anisotropy as a function of frequency between the dynamic and static cases, particularly for transverse polarization.}}
\label{fig:r-fixed-Ld}
\end{figure*}

\begin{figure*}[t] 
\includegraphics[width=0.42\textwidth]{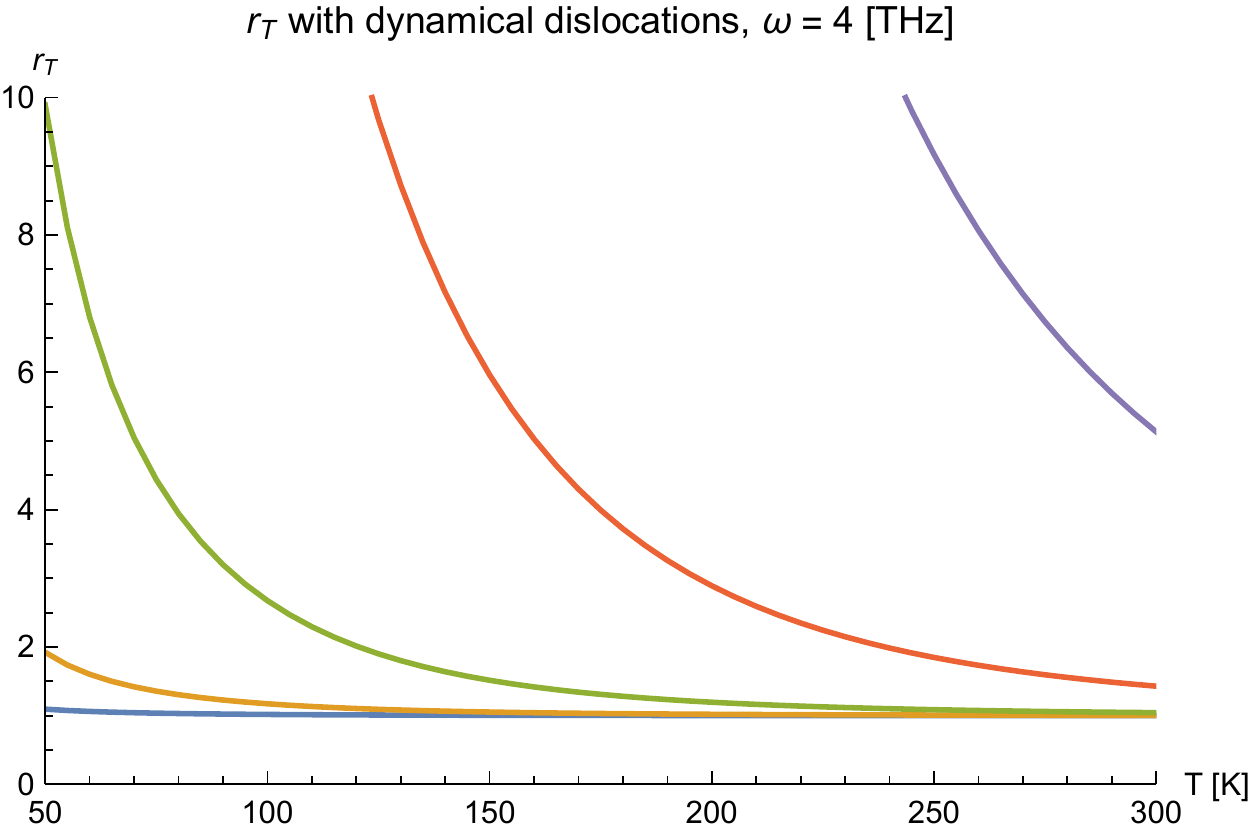}
\includegraphics[width=0.56\textwidth]{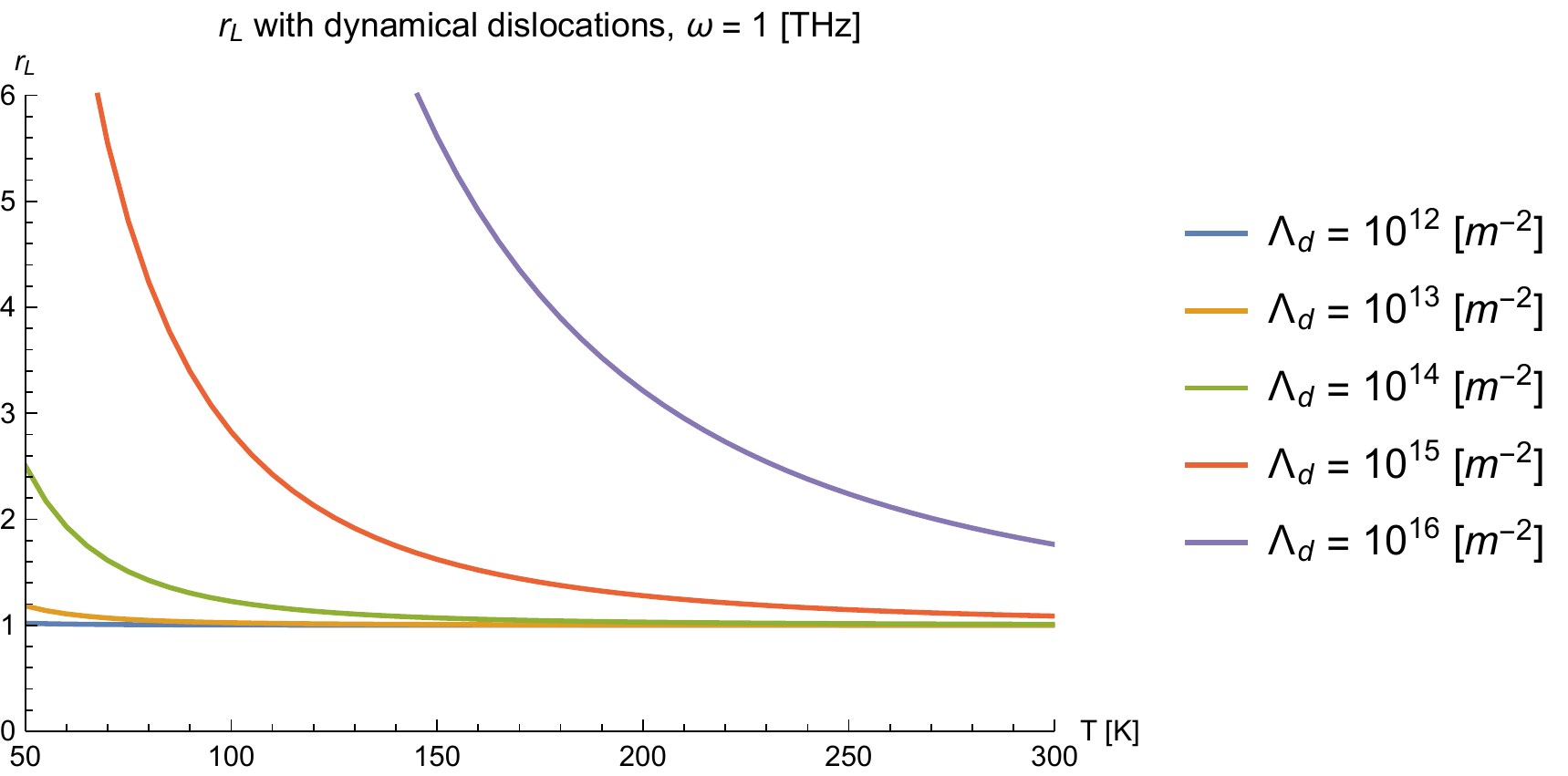}
\includegraphics[width=0.42\textwidth]{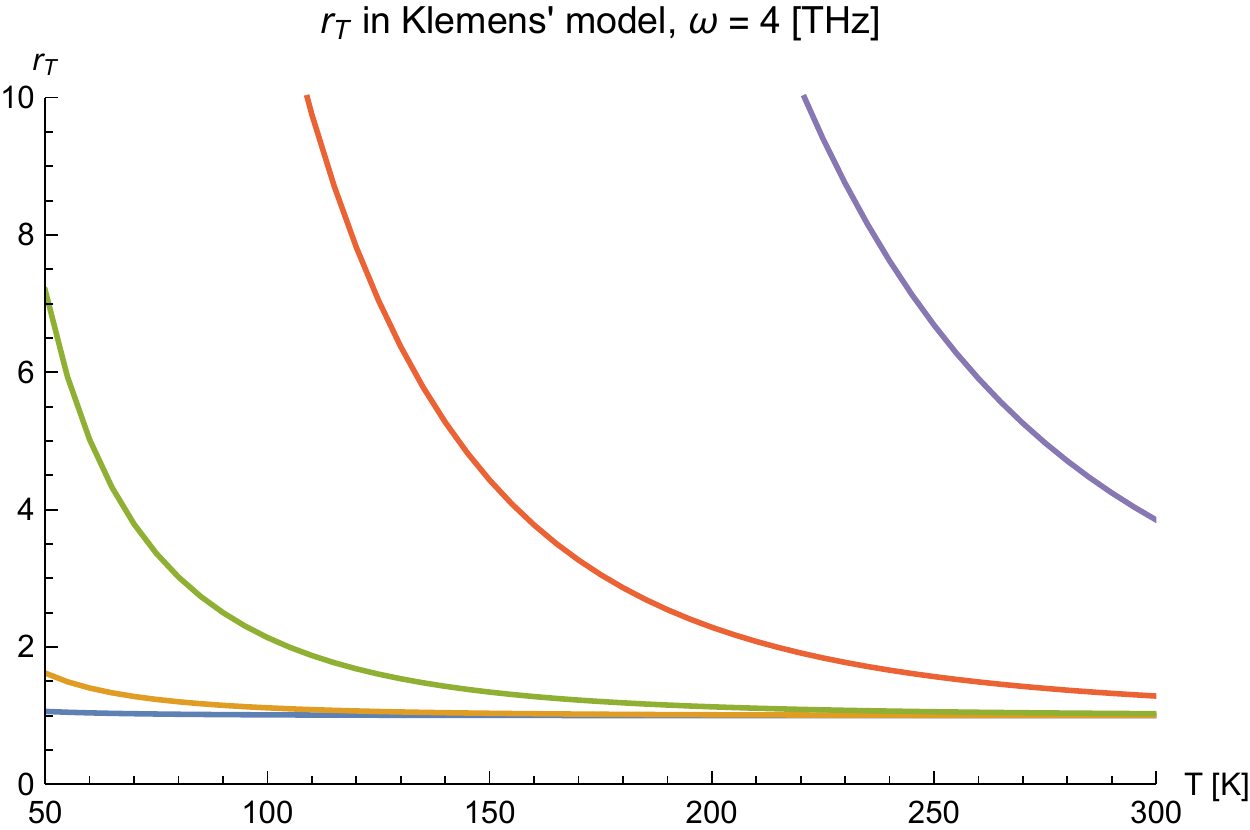}
\includegraphics[width=0.56\textwidth]{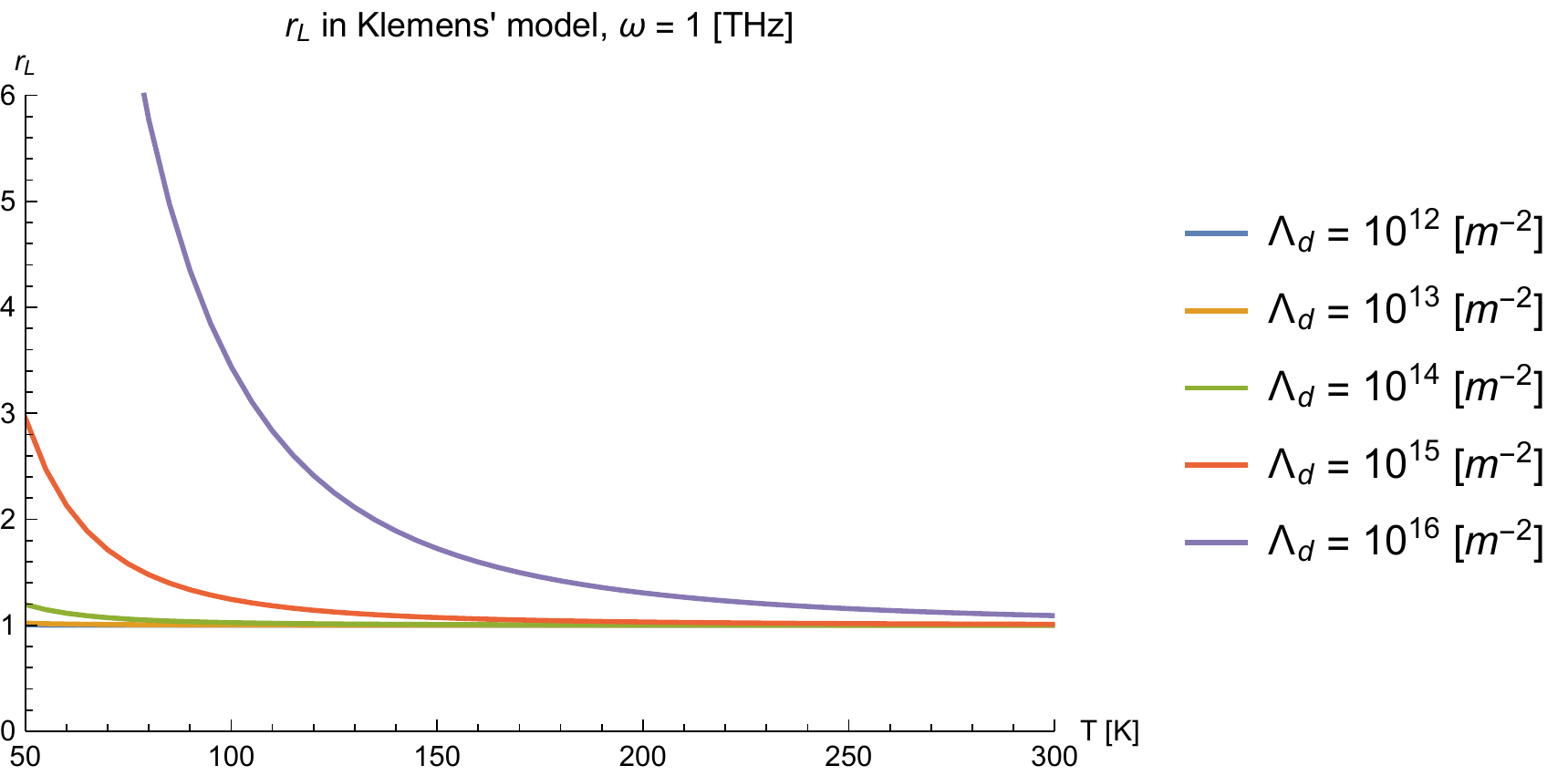}
\caption{{Differential anisotropy ratio $r_\iota$ as a function of temperature $T$ for different dislocation densities $\Lambda_d$. {\it Upper panels:} Present work, based on a dynamical response of dislocations to phonons. {\it Lower panels:} Klemens model, based on a static response of dislocations to phonons. {\it Left panels:} Transverse polarization at $\omega = 4$ THz. {\it Right panels:} Longitudinal polarization at $\omega = 1$ THz. The plots were calculated for  {$\gamma=2$, $g = 3$, $G = 2$, $b = 3.5 \times 10^{-10}$ m,} and the  {intrinsic phonon lifetimes parametrizations} of Ge~\cite{Asen1997}. There is no significant difference between the behavior of the static and dynamic dislocations other than the overall strength of the scattering.}}
\label{fig:r-fixed-w}
\end{figure*}

{{With these definitions in hand, we can now} calculate the anisotropy ratio $r_\iota$ between differential thermal conductivities (per unit frequency/energy $\omega$ and per polarization mode $\iota$) parallel and perpendicular to the dislocation line by writing
\beq \label{r-ratio}
r_\iota \equiv \frac{2 \int_0^\pi d\theta \sin (\theta) \cos^2 (\theta) \tau^\parallel_\iota(\omega, \theta)}{\int_0^\pi d\theta \sin^3 (\theta) \tau^\perp_\iota(\omega, \theta) },
\eeq
{which we shall call a \textit{differential anisotropy ratio---DAR}, as an estimate of how large is the anisotropy in heat transport at each energy scale $\hbar \omega$.}}

{This estimate is most relevant at low temperatures, where intrinsic phonon scattering due to anharmonicities of the elastic continuum becomes subdominant,} {and lends itself to carry out a quantitative comparison between the predictions of our dynamical approach to dislocations and the static approach of Klemens and Carruthers. }

{The first thing to notice is that, in our expressions due to scattering by dislocations,  {$\tau^\perp_T$ and $\tau^\perp_L$ diverge as $\sim 1/\theta^4$ at small polar angles ($\theta \ll 1$), whereas $\tau^\parallel_T$ and $\tau^\parallel_L$ are formally infinite}. This is explicit  {when the temperature gradient is parallel to the dislocations, as both~\eqref{tau-T-par} and~\eqref{tau-L-par} give vanishing inverse lifetimes. When the temperature gradient is perpendicular to the dislocation lines,} one can see from~\eqref{tau-L-per} that $\tau_L^\perp \propto 1/\sin^4(\theta)$, and in $\tau_T^\perp$ one needs to inspect the function $I(\cos (\theta)/c_T)$ to see that an additional factor of $(1 - \cos^4 \theta)$ appears  {in equation~\eqref{tau-T-per}}. This means that, in the absence of other scattering mechanisms, both integrals in~\eqref{r-ratio} are infinite because of the kinematic region where the incident phonon becomes parallel to the dislocation line.} {Roughly speaking, the cross-section for phonon scattering along the dislocation line vanishes. So these phonons proceed unimpeded by dislocations and have an infinite relaxation time. In reality, there are mechanisms, additional to dislocation scattering, that hamper the motion of phonons along the dislocation lines and they must be considered for a realistic assessment.   }

{These mechanisms effectively regulate the aforementioned divergence, and leave the result under quantitative control. Among these mechanisms we highlight that:
\begin{enumerate}
\item there is always ``intrinsic'' phonon scattering due to anharmonicities in the elastic continuum,
\item the dislocation lines will usually not be perfectly aligned in a real material,
\item the consideration of finite size effects in the material introduce a boundary scattering contribution.
\end{enumerate}
In what follows, we will assume that we have a perfectly aligned array of dislocations and we will neglect boundary scattering. Thus, we will only consider intrinsic phonon scattering as the dominant scattering mechanism, besides the scattering by the dislocations themselves.}

{To combine the different decay rates, we use Matthiessen's rule, which in our case means that
\beq
\tau^{{\rm total}}_\iota(\k)^{-1} = \tau_\iota(\omega,\theta)^{-1} + \tau^{\rm intr}_\iota(\omega,T)^{-1},
\eeq
{which is justified as long as the physical processes controlling each lifetime are independent.}
Geometrically, this corresponds to adding the cross-sections of the relevant scattering processes.}

{Now we need estimates for the intrinsic phonon lifetime due to elastic anharmonicities. To get an order of magnitude estimate, we use the following parametrizations~\cite{Asen1997}:
\beq
\tau_T^{\rm intr}(\omega,T)^{-1} = B_T \times \omega T^4 + B_{TU} \times \omega^2 T e^{-C_T/T},
\eeq
\beq
\tau_L^{\rm intr}(\omega,T)^{-1} = B_L \times \omega^2 T^3 + B_{LU} \times \omega^2 T e^{-C_L/T}. 
\eeq
As a working example, we use the values reported by Asen-Palmer et al.~\cite{Asen1997} for Germanium crystals: $B_T = 2 \times 10^{-13} \, {\rm K}^{-4}$, $B_L = 2 \times 10^{-21} \, {\rm s} \cdot {\rm K}^{-3}$, $B_{TU} = 1 \times 10^{-19} \, {\rm s}$, $B_{LU} = 5 \times 10^{-19} \, {\rm s}$, $C_T = 55 \, {\rm K}$, and $C_L = 180 \, {\rm K}$. Also, we use  {$c_T = 3000 \, {\rm m/s}$}.}

{We present results for the differential anisotropy ratios $r_T$ and $r_L$ at various frequencies $\omega$ as a function of temperature $T$ in {Figure~\ref{fig:r-fixed-Ld} (upper panels)}, and at various dislocation densities $\Lambda_d$ in {Figure~\ref{fig:r-fixed-w}.} We chose $\gamma=2$ and $g = 3$ as representative values for the plots. Overall, the anisotropy ratios grow as the temperature of the medium or the frequency of the incident phonons are lowered, and also grow when the dislocation density is increased, as one would qualitatively expect from the form of our phonon lifetimes. We note that the anisotropy ratio is greater for the transverse modes of phonons than for longitudinal polarization; this can be attributed to i) that longitudinally-polarized phonons can scatter with the dislocation even if their angle of incidence is {arbitrarily close to being} parallel to the dislocation {(with the cross section vanishing only in the strict case $\theta = 0$)}, making the anisotropy relatively smaller, and ii) that their decay rate from intrinsic phonon scattering processes is larger, thus needing a larger phonon-dislocation cross-section for this process to be relevant.}

{To compare with Klemens' and Carruthers' models, we note that the inverse phonon lifetimes of both models are linear on the incident phonon energy, and therefore, qualitatively (up to a factor independent of $\omega$), they exhibit the same behavior in the anisotropy ratios. Thus, we take Klemens' model as a point of comparison, taking  {$\tau^{-1} = \omega \Lambda_d b^2 G^2$} for both polarizations. For simplicity, we will also assume that the phonon lifetimes in the presence of a temperature gradient parallel to the dislocation line in this model are negligible.  {This should} provide a conservative benchmark with which to decide whether the model developed herein can explain large anisotropy ratios in thermal conductivities convincingly.}

{We present plots for $r_T$, $r_L$ in Klemens' model for phonon scattering in {Figure~\ref{fig:r-fixed-Ld} (lower panels)}. Comparing with their homologous plots in {the upper panels of Figure}~\ref{fig:r-fixed-Ld}, we see that while the curves are similar for $\omega \sim 1 $ THz, the curves of the anisotropy ratios for other frequencies are much closer to each other in Klemens' model than in ours. This is so precisely because of the different frequency dependence in Klemens' model than in ours: since the phonon decay rate in Carruthers' and Klemens' models is linear in frequency, the anisotropy, which is generated by the difference in relative size between $\tau^{-1}_{\rm dislocation}$ and $\tau^{-1}_{\rm intrinsic}$, is less sensitive to changes in the incident phonon frequency {than in our model} because both decay rates grow with $\omega$. {In contrast to this linear growth in frequency}, in our model the phonon lifetime due to scattering by dislocations decreases as $\omega^{-1}$ with increasing frequency. {Consequently, this makes the differential anisotropy ratio more sensitive to variations in the frequency than in Carruthers' or Klemens' models.}}

{Figure~\ref{fig:r-fixed-w} displays the same differential anisotropy ratios $r_T$ and $r_L$, but this time at fixed frequency and varying dislocation density. Unlike the frequency dependence of the anisotropies, which was bound to be different because of the distinct form of the phonon lifetimes in ours and Klemens' models, their dependence on the dislocation density, illustrated by the distance between the different lines in each plot in Figure~\ref{fig:r-fixed-w} for the two models, is not so different because all of the lifetimes depend linearly on the dislocation density $\Lambda_d$; only the overall strength of the scattering differs.}

{{The above considerations} make our model particularly promising in future attempts to explain large thermal conductivity anisotropies as the temperature is lowered from room temperature to $\sim 50$ K, because at lower temperatures the phonon frequencies/wavenumbers that mainly contribute to the bulk thermal conductivity of an elastic continuum are also smaller. {Correspondingly,} the differential anisotropy ratios $r_\iota$ will grow faster {in the presently considered model} than in Carruthers' or Klemens' models, precisely because the phonon decay rate due to dislocations goes as an inverse power of the frequency instead of linearly.}

\vspace{3mm}

\section{Concluding remarks}
\label{sec:conclusions}
{We have considered a quantum theory of the dynamical modes of an infinitely long dislocation line, modeled as an elastic string, in interaction with phonons, {which are the relevant quantum degrees of freedom at small deformations}, in a continuous, homogeneous, elastic medium. The formalism holds for anisotropic media, and we have presented specific results when the medium is homogeneous. The interaction is through the well-known Peach-Koehler force exerted by a stress on a dislocation line. The quantum interaction depends on a dimensionless coupling constant that depends itself on a short-distance cutoff length at which the continuum theory ceases to be valid,  and the theory is solved to all orders in said constant. Only small excursions of the dislocation line away from its equilibrium position are, however, allowed so that the interaction is quadratic. The behavior of the quanta of dislocation motion (``dislons'') is obtained, and it is revealed that there can be both unstable as well as stable dislons, depending on the strength of the coupling constant. From this information it is possible to estimate the phonon contribution to the internal damping of dislocation motion when they are treated as classical (i.e., non quantum) strings, revealing a linear-in-frequency dependence {for said damping.  {Equivalently, this dissipative term} could be interpreted as a complex contribution to the ``dislon'' sound speed for the modes propagating on the string}. The scattering cross-section for phonons by dislocations is obtained as an explicit function of phonon polarization, angle of incidence and frequency. {In the infinite length approximation we have considered, its dependence on frequency $\omega$ becomes rather simple: it behaves as $\omega^{-1}$}. 
}

{The contribution to the scattering of phonons by dynamic dislocations is considered, especially in comparison with the classical models of phonon scattering by static dislocations of Klemens and Carruthers. In the case of a solid threaded by many parallel dislocations, we consider the ratio between the thermal conductivity per unit frequency for each polarization, in a direction parallel and perpendicular to the dislocation orientation (``differential anisotropy ratio''---DAR), as a function of temperature. Dynamic dislocations yield a DAR that is {considerably more sensitive to frequency than static dislocations}, raising the possibility of a quantitative understanding of recent experimental results on dislocation-induced thermal transport anisotropy {because low-energy phonons are more susceptible to scattering than in previous models~\cite{Klemens1955,Carruthers1959}, and therefore it is possible to have a larger anisotropy at low temperatures.}
}

{We have used a continuum approximation. For the measurements of Sun et al.~\cite{Sun2018}, where the dislocations are one micron in length, this seems a very good approximation. More generally, dislocations typically have lengths in the ten to one hundred nanometer range, where a continuum approach should provide a useful approximation as well. As mentioned in the body of the paper, and implemented explicitely through Eqs. (\ref{eq:dislocmass}-\ref{eq:disloctension}), the theory has only one undetermined dimensionless parameter, the ratio of a long-distance to a short-distance cutoff length. Thus it should be applicable to any crystalline material, irrespective of its microscopic structure, down to length scales of a few interatomic spacings. The other parameters that appear in the formulation we have employed are the mass density and elastic constants, and they are determined from the bulk properties. The Burgers vector, while it appers in the parameters charcaterizing a dislocation, cancels out in the phonon-dislon interaction, as a consequence of this interaction being completely determined by the elastic properties of the material.
}

{We have set up the description of quantum dislocation segments in a quantum field theory framework, which is well suited to include more particles and interactions (such as electrons) in a more complete description of a solid with a large dislocation density. Even though some of the results herein do not depend explicitly on $\hbar$, and therefore could be in principle obtained from an appropriate classical field description, the fundamentally quantum nature of phonons and the length scales involved in forming a dislocation beg for a low-energy quantum-mechanical description, which we have developed through this and earlier work~\cite{Lund2019}. Some purely quantum effects, such as phonon-mediated energy level transitions in a string-like dislocation line are more easily displayed when the dislocation segment is finite and cubic phonon-dislon interactions are considered, although the same transitions are possible in the presently discussed infinite dislocation segments. However, the experimental verification of such features would require a remarkable feat of dislocation engineering in order to be able to isolate the resulting signal and unequivocally attribute a discrete change in the energy of the probe to a specific transition inside the material. A theoretical derivation of a more robust signal that is unequivocally due to the quantum nature of dislocations is also a concrete long-term goal of this description.}

{
A number of possible generalizations of the results presented in this paper suggest themselves:  It should be possible to compute the effect of the third order phonon-dislon interactions, and bring in three-phonon terms. Another direction would be to replace the continuum description with a lattice. Describing phonons in a lattice is standard practice, but the description of dislons, and the corresponding coupling to phonons, would need some care. Also, the interaction with screw dislocations, rather than edge dislocations as carried out in this work, should be straightforward. A specialization, rather than a generalization, would be to consider a two-dimensional lattice, where dislocations are point defects.  This would make their description much simpler and would probably be of relevance for the study of two-dimensional materials \cite{Tan2017,Gu2018,Zeng2020}.
}

Finally, we wish to emphasize that the formalism that has been employed  {in this work}, in conjunction with recent previous results~\cite{Lund2019}, is amenable to {extensions} to include anisotropy, as well as boundary effects, that should make the model suitable for quantitative comparison with experimental data.

\begin{acknowledgments}

We gratefully acknowledge the support of Fondecyt Grant 1191179. Throughout the completion of this work, BSH was supported by a CONICYT grant number CONICYT-PFCHA/Mag\'{i}sterNacional/2018-22181513, by the Thomas Frank fellowship fund at MIT, and by the U.S. Department of Energy, Office of Science, Office of Nuclear Physics under grant Contract Number DE-SC0011090 (Nuclear Theory research).

\end{acknowledgments}

\bibliographystyle{apsrev4-1_title.bst}
\bibliography{QD_bib2.bib}

\end{document}